\documentclass[fleqn,usenatbib]{mnras}

\usepackage{newtxtext,newtxmath}
\usepackage[T1]{fontenc}
\usepackage{ae,aecompl}

\usepackage{graphicx}   % Including figure files
\usepackage{amsmath}    % Advanced maths commands
\usepackage{amssymb}    % Extra maths symbols
\usepackage{mathtools}
\usepackage{lastpage}
\usepackage[utf8]{inputenc}
\usepackage[T1]{fontenc}
\usepackage[portuguese, english]{babel}

\usepackage[usenames, dvipsnames]{color}
\title[The first Super Massive Black Holes]{The first Super Massive Black Holes: indications from models for future observations}

\author[Amarantidis et al.]
{Stergios Amarantidis,$^{1,2}$\thanks{E-mail: samarant@oal.ul.pt}
José Afonso,$^{1,2}$
Hugo Messias,$^{1,3,4}$
Bruno Henriques,$^{5}$
\newauthor Andrew Griffin,$^{6}$
Cedric Lacey,$^{6}$
Claudia del P. Lagos,$^{7,8,9}$
Violeta Gonzalez-Perez,$^{10,11}$
\newauthor Yohan Dubois,$^{12}$
Marta Volonteri,$^{12}$
Israel Matute,$^{1,2}$
Ciro Pappalardo,$^{1,2}$
\newauthor Yuxiang Qin, $^{13}$
Ranga-Ram Chary,$^{14}$
Ray P. Norris$^{15}$
\\ \\
$^{1}$Instituto de Astrofísica e Ciências do Espaço, Universidade de Lisboa, OAL, Tapada da Ajuda, PT1349-018 Lisbon, Portugal\\
$^{2}$Departamento de Física, Faculdade de Ciências, Universidade de Lisboa, Edifício C8, Campo Grande, PT1749-016 Lisbon, Portugal\\
$^{3}$Joint ALMA Observatory, Alonso de Córdova 3107, Vitacura 763-0355, Santiago, Chile\\
$^{4}$European Southern Observatory, Alonso de Córdova 3107, Vitacura, Casilla 19001, 19, Santiago, Chile\\
$^{5}$AV Institute for Astronomy, ETH Zurich, CH-8093 Zurich, Switzerland\\
$^{6}$Institute for Computational Cosmology, University of Durham, South Road, Durham, UK\\
$^{7}$International Centre for Radio Astronomy Research (ICRAR), University of Western Australia, Crawley, Australia\\
$^{8}$ARC Centre of Excellence for All Sky Astrophysics in 3 Dimensions (ASTRO 3D)\\
$^{9}$Cosmic Dawn Center (DAWN), Niels Bohr Institute, University of Copenhagen, Copenhagen, Denmark 0000-0003-3631-7176\\
$^{10}$Institute of Cosmology \& Gravitation, University of Portsmouth, Dennis Sciama Building, Portsmouth, PO1 3FX, UK\\
$^{11}$Energy Lancaster, Lancaster University, Lancaster LA14YB, UK\\
$^{12}$Institut d'Astrophysique de Paris, UPMC et CNRS, Paris, France\\
$^{13}$School of Physics, University of Melbourne, Parkville, VIC 3010, Australia\\
$^{14}$IPAC, Caltech, KS 314-6, Pasadena, CA 91125, USA\\
$^{15}$Western Sydney University, Locked Bag 1797, Penrith South, NSW 1797, Australia}
\date{Accepted XXX. Received YYY; in original form ZZZ}

\pubyear{2019}

\begin{document}

\label{firstpage}
\pagerange{1-17}
\maketitle

\begin{abstract}
We present an exploration of the expected detection of the earliest Active Galactic Nuclei (AGN) in the Universe from state-of-art galaxy formation and evolution semi-analytic models and hydro-dynamical simulations. We estimate the number and radiative characteristics of Super Massive Black Holes (SMBHs) at $z\geq 6$, a redshift range that will be intensively explored by the next generation of telescopes, in particular in the radio through the Square Kilometre Array (SKA) and at high energies with ESA's Athena X-ray Observatory. We find that Athena will be able to observe over 5000 AGN/${\rm deg}^2$ at the Epoch of Re-ionization (EoR), $6\leq z \leq 10$. Similarly, for the same redshift range the models/simulations suggest that SKA will detect at least 400 AGN/${\rm deg}^2$. Additionally, we stress the importance of the volume of the simulation box as well as the initial physical conditions of the models/simulations on their effect on the luminosity functions (LFs) and the creation of the most massive SMBHs that we currently observe at the EoR. Furthermore, following the evolution of the accretion mode of the SMBHs in each model/simulation, we show that, while the quasar dominates over the radio mode at the EoR, detection at radio wavelengths still reaches significant numbers even at the highest redshifts. Finally, we present the effect that the radiative efficiency has on the LFs by comparing results produced with a constant value for the radiative efficiency and more complex calculations based on the spin of each SMBH.
\end{abstract}


%%%%%%%%%%%%%%%%%%%%%%%%%%%%%%%%%%%%%%%%%%%%%%%%%%

%%%%%%%%%%%%%%%%% BODY OF PAPER %%%%%%%%%%%%%%%%%%
\definecolor{col_hor}{RGB}{136,34,85}
\definecolor{col_eag}{RGB}{221,204,119}
\definecolor{col_mii}{RGB}{136,204,238}
\definecolor{col_mun}{RGB}{17,119,51}
\definecolor{col_dur}{RGB}{51,34,136}
\definecolor{col_drag}{RGB}{204,102,119}
\definecolor{col_shar}{RGB}{187,187,187}

\section{Introduction}
One of the fundamental questions in astronomy is how galaxies form and evolve through cosmic time. For the past three decades, various teams have been trying to answer this question by computationally generating a realistic Universe, with more or less detailed physics, and following the birth and evolution of individual simulated galaxies. Two major techniques have been developed and are generally adopted: semi-analytic models (SAMs, \citealt{whiterees}) and hydro-dynamical simulations (HDSs, \citealt{carlberg,katz}). Although there are fundamental limitations to the predictions these models and simulations (models hereafter) can provide, due to computational cost and necessary simplifications for the physical processes involved, the past few years have seen a vast and remarkable improvement in their results when compared to observations.

\indent One of these improvements is the inclusion of a fundamental player in the evolution of a galaxy, an actively accreting Super Massive Black Hole (SMBH) \citep[e.g.][]{benson03,bower2006,croton2006,peng2006,volonteri2007,lagos08,merloni2010,heckman2011}, revealing itself as an Active Galactic Nucleus (AGN). It is believed that most galaxies host a SMBH at their centres \citep[e.g.][and references therein]{kormendy2001}, and that both, galaxy and SMBH, grow somehow in tandem -- therefore studying the formation, growth and feedback of SMBHs is fundamental in understanding the growth of galaxies throughout the Universe's history.

\indent Although relations such as the SMBH-bulge mass can be reproduced by the models in the local Universe \citep[e.g.][]{jahnke,graham15,shirakata16,yang18}, and other observables such as the SMBH mass function \citep[e.g.][]{volonteri16}, stellar mass function \citep[e.g.][]{Kaviraj17} or bolometric luminosity \citep[e.g.][]{griffin18} can be successfully reproduced up to considerable distances ($z \sim 3-4$), the modelling of the first galaxies, at high redshifts ($z \geq 6$), and their SMBHs is still a wide open topic with various possible solutions \citep[e.g.][]{ebisuzaki2001,bromm2003,kousiapas2004,volonteri2010}. Generally, the models place the first seeds of SMBHs (with masses $M_{\bullet} \sim 10^5 \, \rm M_{\odot}$) in halos that exceed a specific mass \citep[e.g.][]{dimateo} or where gas fulfils particular conditions \citep[e.g.][]{yohan12, taylor14, Habouzit}, becoming a fundamental driver, through feedback processes, of galaxy growth from that point onwards. Although this procedure results in the appearance of SMBHs in the very early Universe, it is still unclear if it is sufficient to reproduce the most massive ones currently observed at the highest redshifts \citep[e.g.][]{fan,mortlock,derosa,wu,mazzu,banados,reed}. It has to be mentioned that due to resolution limitations of the models these seeds cannot be resolved. As a result, they are followed through sub-grid recipes \citep[see][for a recent review]{Somerville2015}.

In this context, the fundamental process is the SMBH growth and its link to that of the host galaxy. Three major modes of growth have been proposed theoretically and adopted by the models so far. The first mode, called \textit{quasar} or \textit{radiative mode}, assumes a high accretion rate, which generates a geometrically thin, optically thick disc \citep[][]{Shakura1973}. The accretion leads to strong X-ray emission due to photon up-scattering via inverse-Compton interactions with electrons in the hot corona around the SMBH. The second mode, called \textit{radio} or \textit{jet mode}, occurs at low accretion rates which result from an Advection Dominated Accretion Flow \citep[ADAF -][]{rees1982} that typically produces two bipolar outflows of material (jets) converting the potential energy of the in-falling matter into kinetic energy. These two modes of SMBH growth take place at different levels of accretion rate: if above 1$\%$ of the Eddington accretion limit, quasar mode accretion dominates, radio mode otherwise \citep[e.g.][ and references therein]{fabian12, li2012, heckman14}. A third and final mode occurs on the merging of two galaxies, both containing SMBHs, resulting in a higher mass SMBH with possibly different spin. This phenomenon is well known from the observational point of view as dual-AGN \citep[e.g.][]{dual3, dual4, dual2, dual1}. Although the relevance of this mode depends on the model, it can account for a substantial SMBH growth, as the mass losses in the merger are negligible compared to the previous two modes \citep[e.g.][]{Schnittman,healy17}.

It has been shown that it is necessary to consider AGN feedback in order to improve the predictions for the local SMBH and galaxy mass functions along with other observables \citep[e.g.][]{bower2006,croton2006,lagos08,Hirschmann12}. Although for high redshift the dominant quasar mode plays an important role in galaxy evolution, this feedback comes mostly from the radio mode since it deposits kinetic energy to the surrounding material, possibly stopping the cooling flows that would otherwise lead to an enhancement of the star formation. The parameters that regulate these feedback processes are commonly calibrated \citep[e.g.][]{sijacki07} to match local observations and observed relations (e.g. the $M_{\bullet}-\sigma$ relation). Many studies focused on the AGN feedback, comparing SAMs and HDSs with current observations \citep[e.g.][]{benson2011,Lu2010,Hirschmann2011, fanidakis2013,dubois14, Somerville2015, Guo2015,McAlpine2016}, generally show an acceptable agreement with the observed local SMBH mass function, as well as with the infrared and X-ray Luminosity Functions (XLFs) of AGN.

In this work we explore predictions from 4 SAMs and 4 HDSs for the SMBH/AGN population at high redshifts, within the Epoch of Re-ionization (EoR). In order to achieve this, we first investigate the predictions of the models for the local Universe, comparing with recent observational results in the hard X-ray ($2-10$ keV) and radio (1.4 GHz) regimes \citep[][]{Aird2015,Buchner2015,Miyaji2015,rigby11,smolcic17}. This approach has been presented for some of the models in the past \citep[][]{Fanidakis2011,Khandai2014,Sijacki2014,volonteri16,griffin18} for the X-ray part and for the radio \citep[][]{Fanidakis2011,griffin18}, but typically limited to redshifts below 6 (\citealt{griffin18} extend the X-ray predictions to $z>6$). Subsequently, we examine the predictions from these models to the X-ray and radio emission from AGN at the EoR (redshifts of $6-10$) and we provide estimates for the number of AGN that the next generation of telescopes, namely the Advanced Telescope for High Energy Astrophysics (Athena - \citealt{athena_white}) and the Square Kilometre Array (SKA - \citealt{ska_white}), will observe. We should note that there are other models that have been developed solely with the purpose of exploiting the highest redshift Universe \citep[e.g. the BlueTides HDS,][which can be applied currently to the $7.5-99$ redshift range]{bluetides}. While capable of revealing important results about the earliest Universe, these models lack the comparison to observations, in particular at lower redshifts, a fundamental benchmark to gauge how close model results are to the observable Universe. In this work we only explore models that have been tested against observations at low-to-intermediate redshifts.

This paper is organised as follows: in Section 2 a brief overview of the adopted models is presented, describing their basic features and focusing on details about SMBH parameters. In Section 3, we detail our adopted approach to determine the predicted hard-X-rays and radio LFs, including how some of the involved key parameters were fixed to match local-Universe observables, and how we have established the high-redshift AGN number counts. The results are presented in Section 4, and discussed in Section 5.

\section{Models} \label{models}
Although the aim of both techniques of modelling galaxy evolution is comparable, the differences between SAMs and HDSs are significant, allowing them to be used at some extent as complementary tools. While understanding the internal structure of galaxies within a cosmological context requires the level of resolution included in a HDS, for the exploration of the parameter-space related to larger, statistically meaningful samples one would be more inclined to SAMs. While the scope of this paper is not to review these methods, we briefly mention below some of their basic features necessary to understand their relevance and impact to this work \citep[see][for more detailed reviews]{Somerville2015,Wechsler18}.

In HDSs, evolution of matter is followed by solving the hydrodynamic and gravity equations for the gas, Dark Matter (DM) and stars. The capability to follow particle motions allows, for example, to study in detail the kinematics of a galaxy and the accretion of matter into a SMBH, however at the expense of significant computational power. In order to compensate this cost the volume of the simulations is generally small with typical box sizes of $\sim 100\, \rm Mpc$. It has to be noted that physical processes that occur on scales smaller than the mass resolution (e.g. SMBH accretion, star formation) are modelled using phenomenological `sub-grid' treatments.

For SAMs, on the other hand, the evolution of gas is followed by using analytic approaches \citep[see][for a review]{baugh06}. DM haloes are described by merger trees generated from either N-body DM simulations or using Monte Carlo techniques \citep[e.g.][]{parkinson08}. Physical approximations, which can be significant, may be applied, resulting in a less stringent requirement on computational power. As a result, the volume of SAMs can be significantly larger, reaching box sizes of $\sim 1\, \rm Gpc$.

A comparison between the two methods is beyond the scope of this paper \citep[for such comparisons see for instance][]{benson2001, yoshiba, helly03, saro10, monaco, Guo2015, mitchell18}, even though they address different aspects of galaxy evolution. For a wider exploration of the range of predictions that current state-of-the-art models can provide on the earliest AGN populations, we thus consider both classes of models, adopting in this paper 4 HDSs and 4 SAMs which have been developed and tested over the last few years and are able to match a number of observational indicators at intermediate and low redshifts. A quick description of the most important parameters of each model is presented in Table \ref{param0}. It should be noted that for this work we use the data provided by the teams responsible for each model, without re-running any of the models.

\subsection{Hydro-dynamical models}

Several hydro-dynamical codes have been developed over the last few years, with the two most common approaches of solving the hydro-dynamical equations being the smoothed particle hydrodynamics Particle-Mesh method \citep[SPH; e.g.][GADGET]{springel2001} and the Adaptive Mesh Refinement \citep[AMR; e.g.][RAMSES]{teyssier2002} or its latest formulations (e.g. AREPO - \citealt{springel2010}; GIZMO - \citealt{gizmo}). We adopt 4 recent HDSs that have been successful in predicting several observables, for example, the local SMBH mass function \citep[e.g.][]{Sijacki2014,guevara2016,volonteri16, mutlu18} along with other galactic properties \citep[e.g.][]{dubois14, Vogelsberger14,park18,camps16} as well the effect of AGN feedback on galaxy evolution \citep[e.g.][]{beckmann17, terrazas17}. It should be stressed that several other HDSs exist -- for example, the Nyx code \citep[][]{almgren13}, the MAGNETICUM simulations at different resolutions and sizes \citep[][]{hirschmann14}, the $\nu$2GC simulations \citep[][]{ishiyama15}, the RHAPSODY-G simulations \citep[][]{rhapsody}, the BlueTides simulation \citep[][]{bluetides}, the BAHAMAS simulation \citep[][]{bahamas} or the recent simulation runs from the Illustris team called Illustris-TNG \citep[][]{illustristng}. Also noteworthy is that a significant amount of work has been done on zoomed-in hydro-dynamical simulations (e.g. Aucila comparison project - \citealt{scannapieco12}; FIRE simulation - \citealt{fire}; AGORA simulations - \citealt{kim16}; AURIGA project - \citealt{grand17}; FIRE-2 simulation - \citealt{fire2}; SPHINX simulation - \citealt{SPHINX}; ROMULUSC simulation - \citealt{ROMULUSC}), focusing on the detailed modelling of the evolution of individual galaxies with much higher resolution than the HDSs being used in this paper. The drawback of these simulations is the small volume and, consequently, the low number of galaxies produced, which renders them unusable for the scope of our work.
Therefore, we choose simulations that provide a statistically large sample of galaxies which is translated to a box size of $L \geq 100 \, \rm Mpc$. Although more simulations exist, we have selected 4 based on the accessibility of their data products to the community, considering that the range of their predictions is representative of the overall HDS capabilities.

\subsubsection{Horizon-AGN Simulation}
The Horizon-AGN \citep[][]{dubois14} is a HDS covering a volume of $V = (142\, \rm cMpc)^3$ which makes use of the Adaptive Mesh Refinement code RAMSES \citep[][]{teyssier2002}. The cosmological parameters being used have been derived from the WMAP-7 cosmology \citep[][]{wmap7}, compatible with a Hubble constant of $H_0=70.4\, \rm km s^{-1}Mpc^{-1}$.
By using $1024^3$ DM particles the model achieves a DM mass resolution of $M_{\rm DM,res}=8\times 10^7 \,\rm  M_{\odot}$, baryonic mass resolution of $M_{\rm bar,res}=2\times 10^6\, \rm M_{\odot}$ and spatial resolution of $r_{\rm res}=1 \, \rm kpc$.
The initial SMBH seed is set to $M_{\rm \bullet,seed}=10^5\, \rm M_{\odot}$ being placed in a galaxy when the gas and stellar density exceed the limit of star formation ($0.1\,\rm H \,cm^{-3}$) and the stellar velocity dispersion exceeds the limit of $100 \,\rm km\,s^{-1}$.
The model sets the condition that all SMBHs have been formed by redshift 1.5 and the accretion rate follows a steady, spherically symmetric Bondi-Hoyle-Lyttleton accretion given by $\dot{M}_{\rm Bondi} = \dot{M}_{\bullet}=4 \pi \alpha G^2 M^2_{\bullet} \overline{\rho}/(\overline{c_{\rm s}}^2+\overline{u}^2)^{3/2}$ where $M_{\bullet}$ is the SMBH mass, $\overline{\rho}$ is the average gas density, $\overline{\rm s}$ is the average sound speed, $\overline{u}$ is the average gas velocity relative to the SMBH and $\alpha$ is a dimensional boost factor.
The accretion rate cannot exceed the Eddington accretion limit, therefore rates higher than this value are capped to the Eddington accretion limit ($\dot{m}=\dot{M}_{\bullet}/\dot{M}_{\rm edd} = 1$).
Moreover, the spin parameter of the SMBH is tracked by the model. The SMBH/AGN feedback \citep[][]{yohan12} is provided by two disc modes (quasar and radio mode accretion) separated by $\dot{m}=\dot{M}_{\bullet}/\dot{M}_{\rm edd} = 0.01$, where $\dot{M}_{\rm edd}=L_{\rm edd}/(0.1c^2)$ and $L_{\rm edd}$ is the Eddington luminosity limit.
The radiative efficiency defined as $\epsilon \equiv L_{\rm bol}/(\dot{M}_{\bullet}c^2)$ is set to a constant value of 0.1. The data for the Horizon-AGN simulation can be retrieved through their official website: \href{url}{https://www.horizon-simulation.org/data.html}.

\subsubsection{Illustris Simulation}
The Illustris simulation \citep[][]{illustris_intro} consists of three hydro-dynamical simulation runs of the same volume  $V = (106.5\,\rm cMpc)^3$ and varying resolutions, making use of the AREPO code \citep[][]{arepo}.
In this work we use the simulation with the best resolution, which considers $3\times 1820^3$ DM particles with DM mass resolution of $M_{\rm DM,res}=6.26\times 10^6 \,\rm M_{\odot}$, baryonic mass resolution of $M_{\rm bar,res}=1.26\times 10^6\, \rm M_{\odot}$ and spatial resolution of $r_{\rm res}=0.71\,\rm kpc$.
The cosmological parameters  used have been derived from the WMAP-9 cosmology \citep[][]{wmap9}, compatible with a Hubble constant of $H_0=70.4\, \rm km s^{-1}Mpc^{-1}$. The seed for SMBHs is set to $M_{\rm \bullet,seed} = 1.42\times 10^5 \, \rm M_{\odot}$ in haloes more massive than $M_{\rm halo} = 7.1\times 10^{10} \, \rm M_{\odot}$.
The accretion rate follows the Bondi-Hoyle-Lyttleton described in the previous model and is allowed to exceed the Eddington accretion limit. Rotating SMBHs (and consequently the spin parameter) are not explicitly considered in this model. Three different accretion modes exist for the AGN feedback: the first two correspond to the typical accretion scenarios assumed by most models, while the third mode corresponds to a situation of a disc where the net cooling rate of the gas is modified and accretes close to the Eddington limit. The limit that separates quasar and radio mode is set to $\dot{m}=0.05$ instead of the typical value of 0.01 for calibration reasons and the radiative efficiency on this model was set to $\epsilon = 0.05$ according to \citet{Sijacki2014}. The data can be retrieved from: \href{url}{http://www.illustris-project.org/data/} \citep[][]{nelson15}.

\subsubsection{EAGLE Simulation}
The EAGLE simulation \citep[][]{crain15,schayer15,McAlpine2016} includes 6 hydro-dynamical simulations covering different combination of parameters (e.g. volume, resolution) using the code GADGET-3 (a descendent of the publicly available GADGET-2 code, \citealt{gadget2}). In this work we have used the simulation with the largest volume $V = (100\,\rm cMpc)^3$ and with DM mass resolution of $M_{\rm DM,res}=9.7\times 10^6 \, \rm M_{\odot}$, baryonic mass resolution of $M_{\rm bar,res}=1.81\times 10^6 \, \rm M_{\odot}$ and spatial resolution of $r_{\rm res}=0.7\, \rm kpc$ (these values are valid down to redshift 2.8 where the gravitational softening was fixed - see \citealp[][]{schayer15} for a better description). These simulations are tracking the evolution of baryonic and DM particles with a flat $\rm \Lambda CDM$ cosmology as given by the Planck mission \citep[][]{planck2015}, corresponding to a Hubble constant of $H_0=67.8\,\rm km\,s^{-1}\,Mpc^{-1}$. The SMBH mass seed, of $M_{\rm \bullet,seed} = 1.48\times 10^5 \,\rm  M_{\odot}$, is placed in DM halos more massive than $M_{\rm halo} = 1.4\times 10^{10} \, \rm M_{\odot}$. The accretion rate is described by a modified Bondi-Hoyle-Lyttleton \citep[see][for more details]{guevara2016}: $\dot{M}_{\bullet} = min(\dot{M}_{\rm Bondi}[C^{-1}_{\rm visc}(c_{\rm s}/\overline{u})^3], \dot{M}_{\rm Bondi})$, where $C_{\rm visc}$ is a viscosity parameter.
As in the previous models, the model assumes two disc scenarios separated by $\dot{m} = 0.01$, with the accretion rate capped at the Eddington limit. In this model the spin parameter is not considered, and the AGN feedback is restricted to a single feedback mode which is closest to what we defined as quasar mode \citep[][]{schayer15}. It is assumed that the radiative efficiency is $\epsilon=0.1$, and $1.5 \%$ of the radiated energy is absorbed by the surrounding gas. Data for these simulations can be retrieved through the EAGLE website: \href{url}{http://icc.dur.ac.uk/Eagle/database.php} \citep[][]{McAlpine2016}.

\subsubsection{MassiveBlackII Simulation}
The MassiveBlackII simulation \citep[][]{Khandai2014} is a hydro-dynamical simulation using the P-GADGET code, an updated version of GADGET-2 \citep[][]{gadget2}. The volume of this simulation is $V = (142 \, \rm cMpc)^3$ and the WMAP-7 cosmological parameters with a Hubble constant of $H_0=70.4\,\rm km\,s^{-1}\,Mpc^{-1}$ are assumed. The DM mass, baryonic mass and spatial resolutions are equal to $M_{\rm DM,res}=1.6\times 10^7 \, \rm M_{\odot}$, $M_{\rm bar,res}=3.1\times 10^6 \, \rm M_{\odot}$ and $r_{\rm res}=2.6\, \rm kpc$ respectively, while the initial SMBH seed is set to $M_{\rm \bullet,seed}=7.1 \times 10^5\, \rm M_{\odot}$ and is located in DM halos with a mass larger than the limit $M_{\rm halo}=5 \times 10^{10}\,\rm M_{\odot}$. The accretion rate of the SMBH is set to $\dot{M}_{\bullet}=4 \pi G^2 M^2_{\bullet} \overline{\rho}/(\overline{c_{\rm s}}^2+\overline{u}^2)^{3/2}$ which does not include the boost factor $\alpha$ that was described in the previous models \cite[see][and references therein]{Khandai2014}.  Contrary to the other HDSs considered here, the accretion rate in MassiveBlackII is limited to $2 \, \dot{M}_{\rm edd}$, while the radiative efficiency is kept equal to $\epsilon = 0.1$. The data from this simulation can be retrieved from: \href{url}{http://mbii.phys.cmu.edu/data/}.

\subsection{Semi-analytic models}

The second method to model galaxy formation and evolution was developed before the HDS. SAMs have the advantage of being able to create much larger model universes, using a simplified treatment of some of the physical processes involved \citep[e.g.][for a more detailed description]{baugh06,benson10,Somerville2015}. With the initial thought from \citet{whiterees} and later work \citet{cole91}, \citet{lacey91}, \citet{white91}
SAMs became a well established methodology for simulating the Universe. In this paper 4 SAMs have been considered, namely L-Galaxies, GALFORM, MERAXES and SHARK. Other models do also exist, including GAMETE \citep[][]{gamete}, \citet{somer08}, the GALACTICUS project \citep[][]{galacticus}, the eGalICS model \citep[][]{eGalICS}, SAGE \citep[][]{sage}, the Gaea model \citep[][]{gaea} and SAG \citep[][]{sag,sag18}.
Previous studies had compared several of these SAMs \citep[e.g.][]{lu14,samcompar} presenting predictions for the stellar mass function, star formation rate histories and SMBH-bulge mass relation for $z<6$.

\subsubsection{L-Galaxies Model}
The L-Galaxies model (\citealp[][]{hen2015}, based on the model by \citealp[][]{guo11}), also often known as Munich model in the literature, is a SAM which is built and follows the evolution of the DM trees from the Millennium simulations \citep[][]{milenium}.
Here we consider the simulation by \citet{hen2015} with volume of $V = (714\,\rm cMpc)^3$, DM mass resolution of $M_{\rm DM,res} = 1.43\times 10^9 \, \rm M_{\odot}$ and spatial resolution of $r_{\rm res}=5\,\rm kpc$, with cosmological parameters matching the Planck's first year data  \citep[see][for a complete description]{hen2015}, corresponding to a Hubble constant of $H_0=67.3\,\rm  km\,s^{-1}\,Mpc^{-1}$. While the spin of the SMBH is not considered in this model the accretion is similar to the aforementioned models considering quasar and radio modes and capped to the Eddington accretion limit.
The main growth channel of the SMBHs in this model happens during the galaxy merger phase, with the increase in mass described by: $\Delta M_{\bullet}=f_{\bullet}(M_{\rm sat}/M_{\rm cen})M_{\rm cold}/(1+(V_{\bullet}/V_{\rm 200c})^2)$, where $M_{\rm sat}$ and $M_{\rm cen}$ are the masses of the satellite and central merged galaxies, $M_{\rm cold}$ is their total cold mass, $V_{200c}$ is the virial velocity of the DM halo and $f_{\bullet}$, $V_{\bullet}$ are adjustable parameters. The AGN feedback is provided by the radio mode in terms of relativistic jets, with the energy output from the SMBH to the ISM equal to 10$\%$ of the accreted mass. The data can be retrieved from the website: \href{url}{http://gavo.mpa-garching.mpg.de/portal/}.

\subsubsection{GALFORM Model}
In this work we use a version of the GALFORM model \citep[][]{cole2000,lacey16} which implements an improved treatment of the growth of black holes \citep[][]{griffin18}. The model follows the Millennium N-body DM simulation with volume of $V = (800\,\rm cMpc)^3$ usually referred in the literature as P-Millennium \citep[e.g.][]{baugh18,cowley18} using a Planck cosmology corresponding to a Hubble constant of $H_0=67.8\,\rm km\,s^{-1}\,Mpc^{-1}$ \citep[][]{planck13}. The
DM mass resolution is equal to $M_{\rm DM,res} = 1.6\times 10^8 \, \rm M_{\odot}$, the spatial resolution is equal to $r_{\rm res}=3.4\,\rm kpc$, while the SMBH seed mass is set to $M_{\rm \bullet,seed} = 10\,\rm h^{-1} M_{\odot} = 14.8 \, \rm M_{\odot}$ introduced in every halo independent of mass. The accretion of matter onto the SMBH takes place by accretion of gas during starbursts triggered either by mergers or disc instabilities (quasar mode), accretion of gas from the halo’s hot atmosphere (radio mode) and by SMBH merging. The radio mode includes a prescription for AGN feedback in which heating by the SMBH balances gas cooling in haloes while SMBH mergers contribute significantly to the growth of the SMBH \cite[][]{griffin18} especially for $M_{\bullet}>10^8 \, \rm M_{\odot}$. This accretion of gas (which is not capped to the Eddington limit) transfers angular momentum to the SMBH causing changes to its spin which are tracked in the model and this is used in calculating a spin-dependent radiative efficiency. In this paper we are using a different bolometric correction and obscuration fraction, to those used in \citep[][]{griffin18}. The data from this model can be retrieved by contacting the GALFORM team through \href{url}{http://galaxy-catalogue.dur.ac.uk}.

\subsubsection{MERAXES Model}
The MERAXES model \citep[][]{dragons,qin17} is part of the Dark-ages, Re-ionization And Galaxy-formation Observables Numerical Simulation project (DRAGONS) focusing in modelling the EoR. From the two existing DM simulation boxes that DRAGONS is build on we choose the largest volume of $V = (184\,\rm cMpc)^3$ with a DM mass resolution of $M_{\rm DM,res} = 1.2 \times 10^8 \,\rm M_{\odot}$, spatial resolution of $r_{\rm res}=3.4\,\rm kpc$ and following the latest Planck cosmology corresponding to a Hubble constant of $H_0=67.8\, \rm km\,s^{-1}\,Mpc^{-1}$. The SMBH seed mass is set to $M_{\rm \bullet,seed} = 1476 \, \rm M_{\odot}$ and is placed in every newly formed galaxy. The model adopts a Bondi-Hoyle accretion model proposed in \citet{sage} and follows the standard two accretion modes as well as the feedback process described above by setting the value of the radiative efficiency equal to 0.06 (opposed to the typical value of 0.1). The spin of the SMBH is not provided in this model, and the accretion is limited to the Eddington limit. The data can be retrieved by contacting the team through: \href{curl}{http://dragons.ph.unimelb.edu.au}.

\subsubsection{SHARK Model}
The SHARK model \citep[][]{shark}, is a new, flexible, publicly available SAM which is built upon the DM halo catalogs and trees of the SURFS N-body simulations suite \citep[][]{surf}. Here we consider the SURFS simulation with volume $V = (310 \,\rm cMpc)^3$, DM mass resolution of $M_{\rm DM,res} = 3.26 \times 10^8 \,\rm M_{\odot}$ and spatial resolution $r_{\rm res} = 6.64\, \rm kpc$, with cosmological parameters matching the \citet{planck2015}, corresponding to a Hubble constant of $H_0 = 67.51 \rm \,km\, s^{-1}\, M pc^{-1}$. SHARK seeds all halos of masses $>10^{10} h^{-1}\,\rm M_{\odot}$ with SMBHs of masses $10^4 h^{-1} = 14749\,\rm M_{\odot}$. As the other SAMs, Shark considers three channels for the growth of SMBHs: BH-BH mergers, quasar and radio modes. No Eddington accretion limit is imposed. The main growth channel of the SMBHs in this model is starbursts, which are driven by galaxy mergers and disk instabilities (with the two processes playing a similar role in the growth of SMBHs). The increase of mass during starbursts is described by: $\Delta\,\rm M_{\bullet} = f _{\bullet}\,M_{\rm cold} /(1 + (V /V_{\rm vir})^2)$, where $M_{\rm cold}$ is the total interstellar medium mass available for the central starburst, $V_{\rm vir}$ is the virial velocity of the DM halo and $f_{\bullet}$ and $V_{\bullet}$ are adjustable parameters. Feedback from AGN is provided by the radio mode in terms of relativistic jets, with the energy output from the SMBH used to directly reduce or completely quench the cooling flow. Shark, does not follow the spin development of SMBHs, and adopts a fixed radiation efficiency of $0.1$. The model can be retrieved from the website:  \href{curl}{https://github.com/ICRAR/shark} and the data can be accessed by contacting the team at surfs@icrar.org.

\begin{table*}
\centering
\caption{Description of important parameters concerning the basic features of the models and their SMBH formation and evolution.}
\label{param0}
\begin{tabular}{c|c|c|c|c|c|c|c|c}
\hline
Model  & $^{(1)}$Type & $^{(2)}$Size [cMpc] & $^{(3)} M_{\rm \bullet,seed}\,\, [\rm M_{\odot}]$ & $^{(4)} M_{\rm DM,res} \,\, [\rm M_{\odot}]$& $^{(5)} r_{\rm res} \,\, \rm [kpc]$  & $^{(6)} \epsilon$ & $^{(7)} $spin & $^{(8)} \dot{M}_{\bullet}/\dot{M}_{\rm edd}$ \\ \hline
\textcolor{col_hor}{Horizon-AGN} & HDS        &   $142$         &   $10^5$ & $8.0\times 10^7$ &  1.0  & 0.1  & provided & $\leq 1$    \\
\textcolor{black}{Illustris} & HDS     &      $107$       &     $1.42\times10^{5}$ & $6.3\times 10^6$ &  0.7 & 0.05  & N/A  & SE   \\
\textcolor{col_eag}{EAGLE} & HDS        &       $100$       &     $1.48\times10^{5}$ & $9.7\times 10^6$ &  0.7  & 0.1  & N/A     & $\leq 1$  \\
\textcolor{col_mii}{MassiveBlackII}& HDS &     $142$       &     $5\times10^{5}$ & $1.6\times 10^7$ &  2.6  & 0.1  & N/A  & $\leq 2$\\
\textcolor{col_mun}{L-Galaxies}& SAM     &    $714$       &     742 & $1.4\times 10^9$ &  5.0   & 0.1  & N/A  & $\leq 1$    \\

\textcolor{col_dur}{GALFORM} & SAM       &      $800$       &   14.8 & $1.6\times 10^8$ &  3.4   & varies  & provided   & SE   \\
\textcolor{col_drag}{MERAXES}& SAM     &   $184$       &   1476 & $1.2\times 10^8$ &  3.4   & 0.06  & N/A  & $\leq 1$    \\
\textcolor{col_shar}{SHARK}& SAM     &   $310$       &   14749 & $3.4\times 10^8$ &  6.6   & 0.1  & N/A  & SE    \\ \hline
\end{tabular}\\
\scriptsize $^{(1)}$ Model method (SAM: Semi-Analytic Model, HDS: Hydro-Dynamical Simulation), $^{(2)}$ the comoving box size of the simulated Universe in units of Mpc, $^{(3)}$ the SMBH mass seed in units of solar mass, $^{(4)}$ the Dark Matter mass resolution in units of solar mass, $^{(5)}$ the spatial resolution in units of kpc, $^{(6)}$ the radiative efficiency of converting matter into radiation, $^{(7)}$ the spin of the SMBH (provided: the model follows the evolution of the SMBH spin, N/A: the spin is not provided), $^{(8)}$ the accretion rate in units of Eddington accretion ($\leq 1,2$: accretion is capped to this value, SE: no limits in the accretion).
\end{table*}

\section{Methodology} \label{methods}
The models described above are up-to-date simulations of galaxy formation and evolution, frequently being tested against observations and, consequently, updated or improved. They can thus provide estimates of the highest redshift ($z>6$) Universe, in a way that is arguably much more powerful (or at least better physically justified) than extrapolating from lower-redshift observations assuming some undetermined LF evolution towards the highest redshifts.

A fundamental step is the conversion from physical parameters in the model, like mass, spin, and accretion rate, to observables, namely luminosity. In this section we present the relevant steps implemented in this work in order to obtain the luminosity in two essential wavelength regimes for the observation of high redshift AGN: X-rays and radio. In particular, we focus on the hard X-ray ($2-10$ keV) and radio (1.4 GHz) regimes, and aim to extract, from the models, estimates for the hard X-ray and radio luminosity functions (HXLFs and RLFs hereafter).

A particularly relevant point is the rotation of SMBHs (considered in some of the models as the spin parameter - $a$), which will determine the amount of infall matter that will be converted into radiation (i.e. radiative efficiency - $\epsilon$). Six of the models considered in this work (Illustris, EAGLE, MassiveBlackII, L-Galaxies, MERAXES and SHARK) do not track such information, in which case the usual practice is to assume a constant value for the radiative efficiency. A value of $\epsilon=0.1$ equivalent to $a=0.67$ \citep[see equation 2.21 in][for the calculation of $a$]{Bardeen1972} is commonly used in the models, in order to reproduce the observational properties of galaxies in the local Universe \citep[e.g.][]{Khandai2014,guevara2016,volonteri16}, meaning that, on average, 90$\%$ of the infalling matter will actually accrete into a SMBH contributing to its growth, while 10$\%$ is converted into radiation. In this work, while the constant value of 0.1 has been used (for Illustris we use 0.05 corresponding to a spin value equal to -0.26 and for MERAXES 0.06 corresponding to a spin of 0.083), we have also explored the impact of considering a variable value for the radiative efficiency $\epsilon$ (and spin parameter $a$) whenever the spin parameter is available from the models and whenever possible (GALFORM). A more detailed discussion of the results and possible implications for future observations is presented in Section \ref{results}.

\subsection{X-ray Luminosity functions}\label{xraylfs}
For the calculation of the HXLFs the bolometric luminosity ($L_{\rm bol}$) was estimated for each SMBH, and converted to X-ray luminosity as detailed below.

To estimate the bolometric luminosity due to accretion to a SMBH, two cases need to be distinguished: the quasar accretion mode (Thin Disk scenario - TD) and the radio accretion mode (ADAF - which is assumed to take place whenever the accretion rate is below 1$\%$ of the Eddington accretion limit $\dot{m} < 0.01$).
For the former, the bolometric luminosity is simply given as $L^{\rm TD}_{\rm bol} = \epsilon \dot{M}_{\bullet} c^2$ where $\epsilon$ is the radiative efficiency, $\dot{M}_{\bullet}$ the accretion rate of matter into the SMBH and $c$ is the speed of light \citep[][]{Shakura1973}. A detailed description of the calculation of $\epsilon$ is presented in \ref{radeff}. For the radio accretion mode the calculation of the bolometric luminosity ($L^{\rm ADAF}_{\rm bol}$) is more complex, and will depend on how the accretion compares with the accretion level threshold ($\dot{m}_{\rm crit,\nu}$), which marks the electron-heating being dominated by viscous or ion-electron-heating \citep[][]{mahadevan97}. Here, we follow the equations for AGN bolometric luminosity from \citet{griffin18}, covering both TD and ADAF scenarios:

\begin{equation}
L_{\rm bol}=\left
  \{\begin{array}{l}
  \text{[if $\dot{m} < \dot{m}_{\rm crit,\nu}$]:}\\
    0.0002 L^{\rm TD}_{\rm bol}\bigg(\frac{\delta}{0.0005}\bigg)\bigg(\frac{1-\beta}{0.5}\bigg)\bigg(\frac{6}{\hat{r}_{\rm lso}}\bigg),\\\\
\text{[if $\dot{m}_{\rm crit,\nu} \leq \dot{m} < 0.01$]:}\\
    0.2 L^{\rm TD}_{\rm bol}\bigg(\frac{\dot{m}}{\alpha^2_{\rm ADAF}}\bigg)\bigg(\frac{\beta}{0.5}\bigg)\bigg(\frac{6}{\hat{r}_{\rm lso}}\bigg),
 \quad \\\\
 \text{[if $0.01 \leq \dot{m} < \eta_{\rm edd}$]:}\\
    L^{\rm TD}_{\rm bol},\\\\
\text{[if $\dot{m} \geq \eta_{\rm edd}$]:}\\
   \eta_{\rm edd}(1+ln(\dot{m}/\eta_{\rm edd}))L_{\rm edd}
  \end{array} \right.
 \label{equation1}
\end{equation}

\noindent where:
\begin{equation}
\dot{m}_{\rm crit,\nu} = 0.001 \bigg(\frac{\delta}{0.0005}\bigg)\bigg(\frac{1-\beta}{\beta}\bigg)\alpha^2_{\rm ADAF},
\end{equation}

\noindent is the aforementioned boundary, with $\delta$ being the fraction of the viscously dissipated energy received by electrons in an accretion flow (set here to 0.0005), $\alpha_{\rm ADAF}$ is the Shakura-Sunyaev viscosity parameter for the ADAF case (taken here as 0.1), $\beta$ is the ratio of gas pressure to total pressure related to $\alpha_{\rm ADAF}$ by $\beta=1-\alpha_{\rm ADAF}/0.55$ and the parameter $\eta_{\rm edd}$ is a free parameter set equal to 4. A comparison between the resulting bolometric luminosity calculated with equation (\ref{equation1}) 
and the simple version ($L_{\rm bol} = \epsilon \dot{M}_{\bullet} c^2$) demonstrates significant differences 
for $L_{\rm bol}<10^{45}\, \rm erg/s$, however smaller differences (below $\sim 30 \%$) 
for the high end of the LF, with the latter being the relevant range affecting our final predictions.

The X-ray luminosity from accretion to SMBHs can now be estimated from the bolometric luminosities using the corrections \citep{Hopkins2007}:
\begin{gather}
L_{\rm X-ray(2-10 keV)} = \frac{L_{\rm bol}}{10.83 \, (\frac{L_{\rm bol}}{10^{10}\rm L_{\odot}})^{0.28} + 6.08 \, (\frac{L_{\rm bol}}{10^{10}\rm L_{\odot}})^{-0.02}}.
\label{hopkins}
\end{gather}
These corrections are valid for bolometric luminosities of $L_{\rm bol} \sim 10^{41}-10^{49} \, \rm erg/s$ and redshift range of $z=0-6$.
Although studies calculating bolometric corrections use large sets of observed quasar catalogues, they are limited in redshift, which renders the study of the EoR more complicated. In this work we use equation \ref{hopkins} as valid for $z>6$ since we lack observations to determine the corrections for the EoR. Figure \ref{hxrlfs} (left panels) presents the resulting HXLFs for low and intermediate redshifts, following the procedures described above.

\subsection{Radio Luminosity functions}\label{radiolfs}
\begin{figure*}
\centering
\includegraphics[width=\columnwidth]{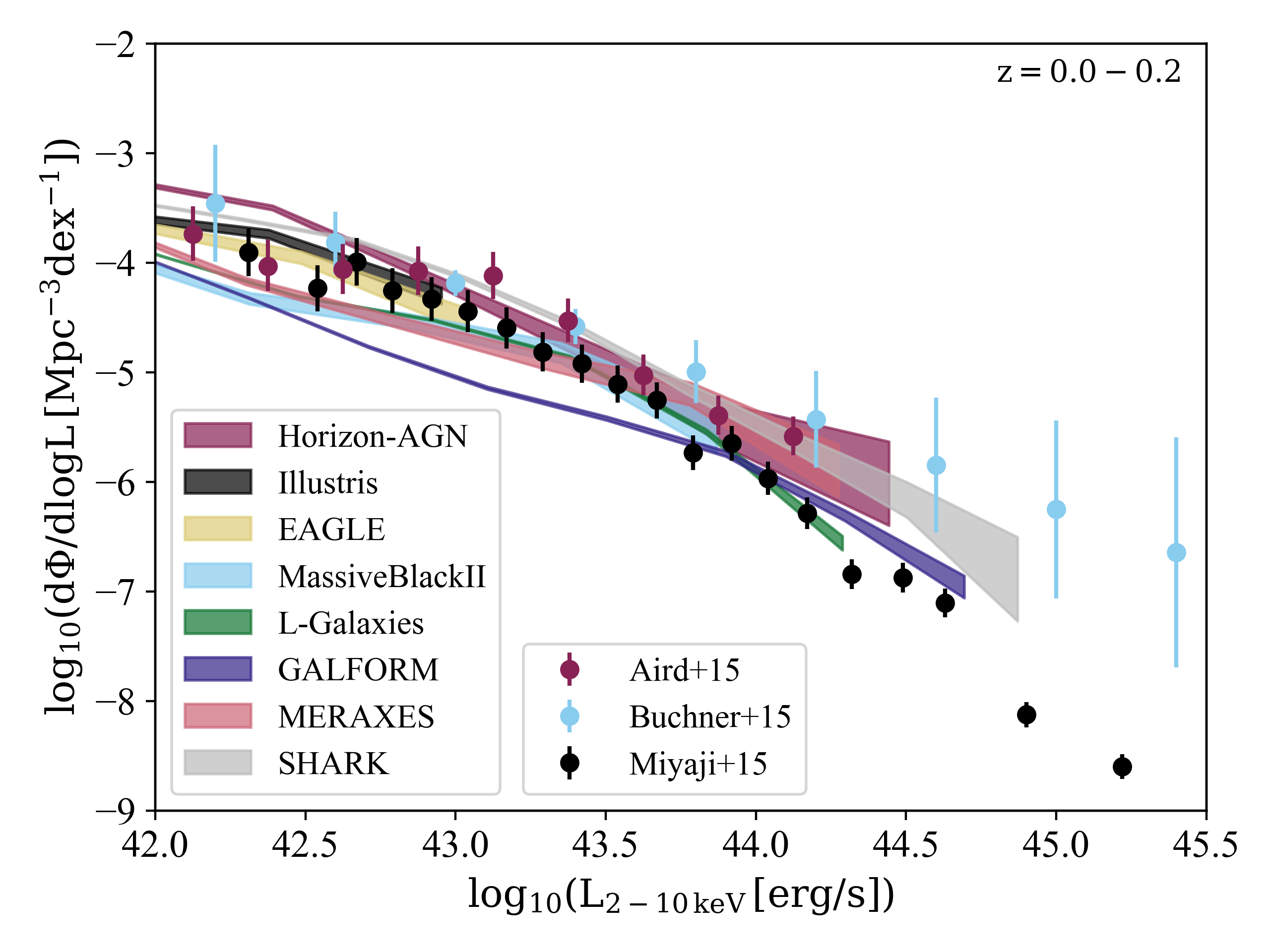}
\includegraphics[width=\columnwidth]{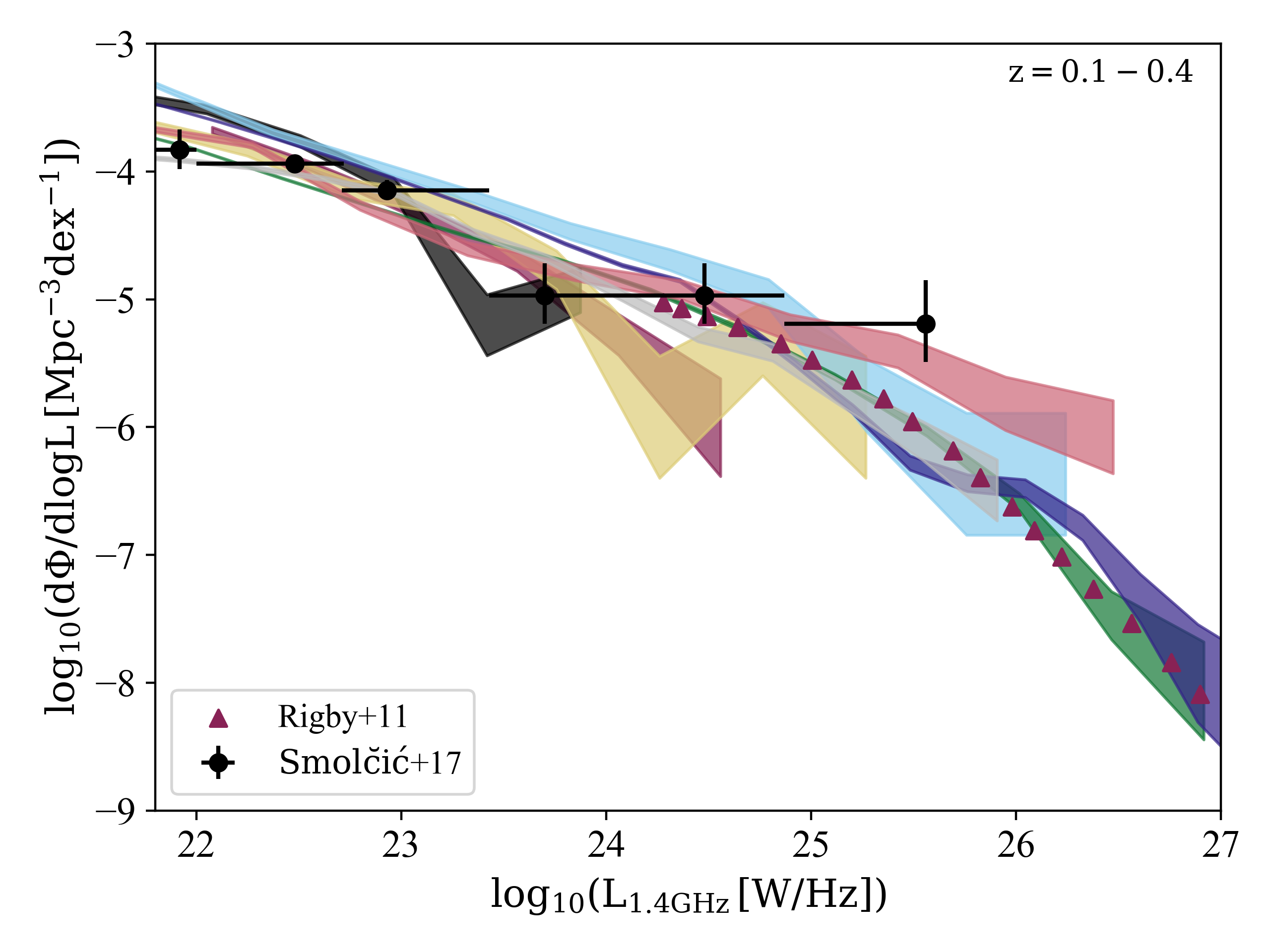}
\includegraphics[width=\columnwidth]{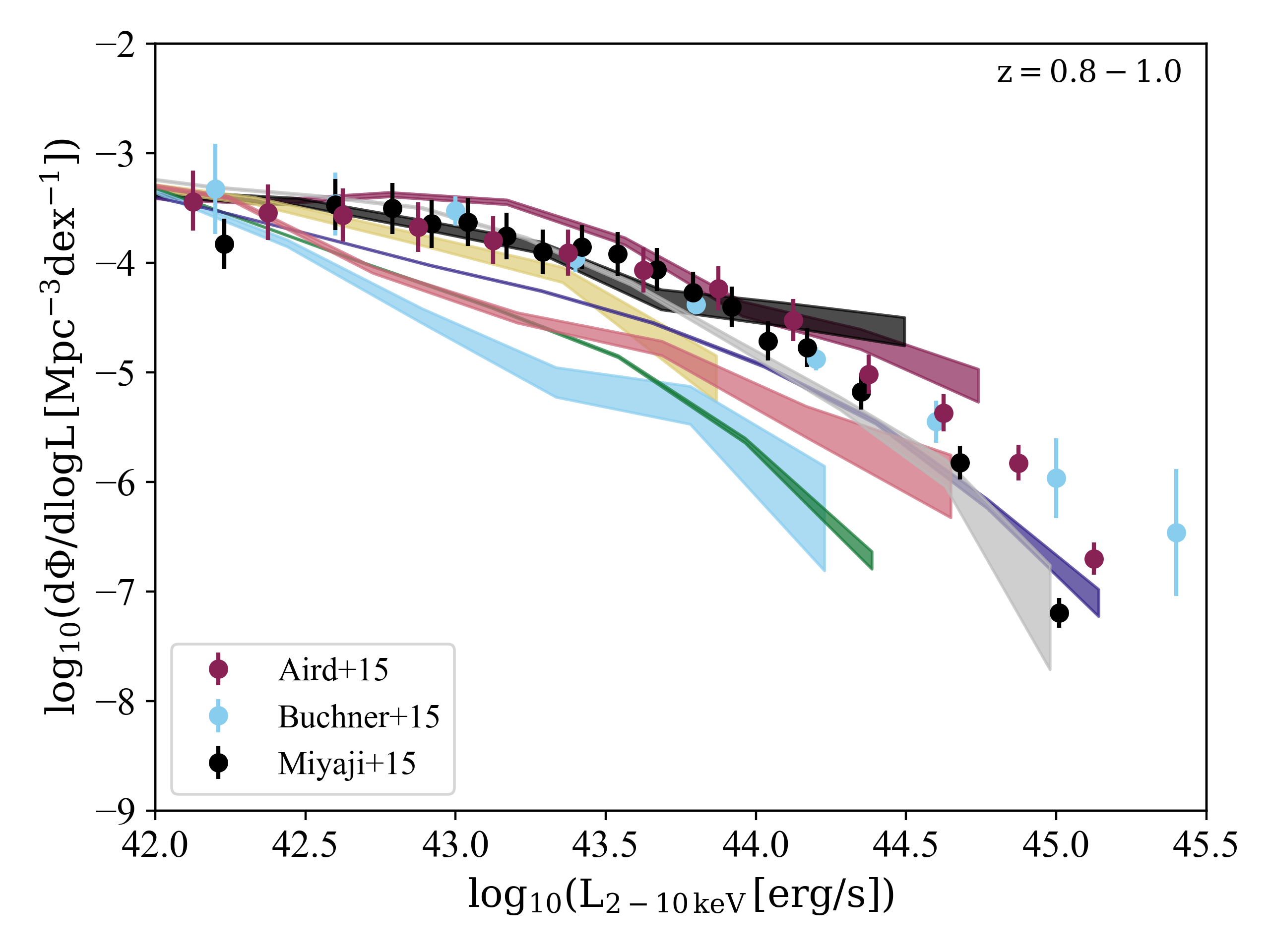}
\includegraphics[width=\columnwidth]{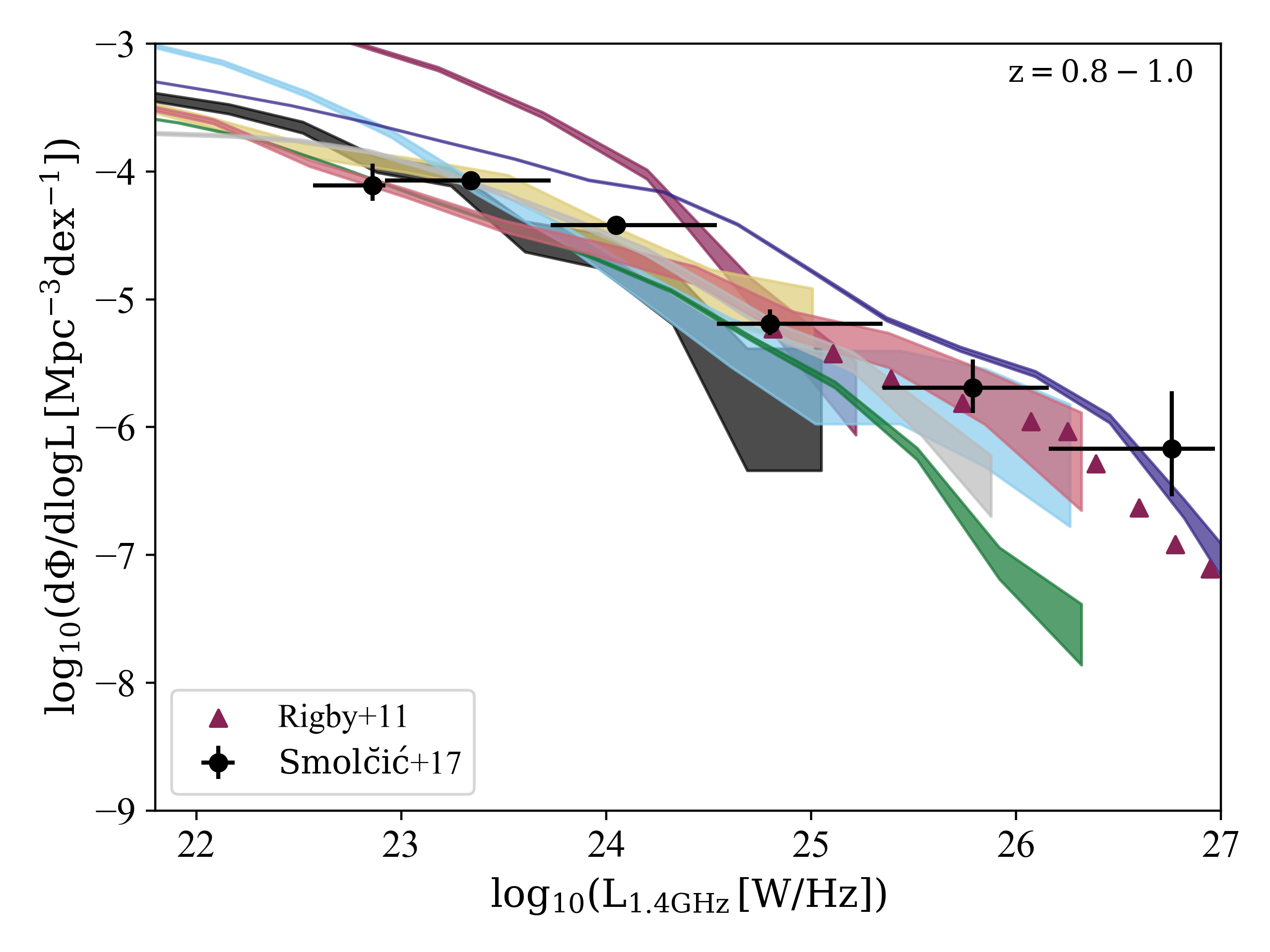}
\caption{The hard X-ray ($2-10$ keV) luminosity functions (left panels) and the radio (1.4 GHz) luminosity functions (right panels) for the local Universe (upper panels) and the redshift range $0.8<z<1.0$ (bottom panels). The thickness of each line corresponds to the Poisson statistical errors. A constant value for the radiative efficiency was applied for all models except for the GALFORM SAM where a spin dependent efficiency was used (see \ref{radeff} for details). For the Horizon-AGN and Illustris models the mass limits of $M_{\rm halo}>5\times 10^{11} \, \rm M_{\odot}$ and $M_{\bullet}>5\times 10^{7}\, \rm M_{\odot}$ were applied respectively, according to \citet{volonteri16} and \citet{Sijacki2014}. The points denote observational results from \citet{Aird2015}, \citet{Buchner2015}, \citet{Miyaji2015}, for the X-rays and from \citet{rigby11}, \citet{smolcic17} for the radio.}
\label{hxrlfs}
\end{figure*}

The determination of the RLF considers the radio emission arising from both the quasar (TD) and the radio (ADAF) accretion modes. We follow the procedures adopted in previous works \citep[e.g.][and references therein]{meier02,Fanidakis2011,2017arXiv171207129I}:
\begin{gather}
\nu L^{\rm ADAF}_{\nu} = A_{\rm ADAF} \bigg(\frac{M_{\bullet}}{10^{9}\rm M_{\odot}} \times \frac{\dot{m}}{0.01}\bigg)^{0.42} L^{\rm ADAF}_{\rm jet}, \\
\nu L^{\rm TD}_{\nu} = A_{\rm TD} \bigg(\frac{M_{\bullet}}{10^{9}\rm M_{\odot}}\bigg)^{0.32}  \bigg(\frac{\dot{m}}{0.01}\bigg)^{-1.2} L^{\rm TD}_{\rm jet},
\end{gather}
where $\nu$ is the radio frequency, $L_{\rm \nu}$ is the radio luminosity density in W/Hz, $A_{\rm ADAF}$ and $A_{\rm TD}$ are normalization factors, and the luminosities of the jets for each mode, are given by:
\begin{gather}
L^{\rm ADAF}_{\rm jet} = 2 \times 10^{45} \bigg(\frac{M_{\bullet}}{10^9 \rm M_{\odot}}\bigg)\bigg(\frac{\dot{m}}{0.01}\bigg) \, a^2  \, \,\rm  [erg/s],\\
L^{\rm TD}_{\rm jet} = 2.5 \times 10^{43} \bigg(\frac{M_{\bullet}}{10^9 \rm M_{\odot}}\bigg)^{1.1}\bigg(\frac{\dot{m}}{0.01}\bigg)^{1.2}a^2 \, \, \rm [erg/s].
\end{gather}
As far as the TD scenario is concerned the combination of equations 5 and 7 indicates that the total radio luminosity from this mode depends on the SMBH mass and spin only, since the $\dot{m}$ terms in both equations cancel out (see Appendix \ref{appB} for more details). It is noteworthy that the values of the normalization parameters, $A_{\rm ADAF}$ and $A_{\rm TD}$, can be significantly different for different models in order to match the local LFs. For example, the GALFORM SAM has changed these parameters quite substantially over time, as modifications in the model and different observational constraints were used.
Since the models are so sensitive to these parameters and it is important to ensure a high degree of consistency in any comparison between them, we normalise the $A_{\rm ADAF}$ and $A_{\rm TD}$ values in all models in order to fit the same observed local RLFs \citep[][]{rigby11,smolcic17}. Table \ref{param} and Figure \ref{hxrlfs} (right panels) show the result of this exercise. While the former shows how different the values for the normalization parameters are, it is important to underscore the existence of a large degeneracy between $A_{\rm ADAF}$ and $A_{\rm TD}$. In order to solve this issue we are choosing the combination of parameters that provide the best fitting to the local and higher z RLFs (see Appendix \ref{appA} for more details).

\begin{table}
\centering
\caption{The values of the two normalization parameters, $A_{\rm ADAF}$ and $A_{\rm TD}$, for the calculation of the radio luminosity for each model.}
\label{param}
\begin{tabular}{c|c|c}
\hline
Model          & $A_{\rm ADAF}$ & $A_{\rm TD}$ \\ \hline
\textcolor{col_hor}{Horizon-AGN}        &   $1.3\times10^{-9}$         &   $3.0\times10^{-3}$        \\
\textcolor{black}{Illustris}      &      $7.0\times10^{-6}$       &     $1.0\times10^{-2}$      \\
\textcolor{col_eag}{EAGLE}         &        $8.0\times10^{-6}$     &     $3.0\times 10^{-2}$      \\
\textcolor{col_mii}{MassiveBlackII} &     $1.5\times10^{-4}$        &       $1.0\times10^{-2}$    \\
\textcolor{col_mun}{L-Galaxies}     &    $8.0\times10^{-5}$         &     $5.0\times10^{-2}$      \\
\textcolor{col_dur}{GALFORM}        &      $2.0\times 10^{-5}$      &    $8.0\times 10^{-1}$      \\
\textcolor{col_drag}{MERAXES}     &   $8.0\times 10^{-5}$      &   $5.0\times 10^{-2}$      \\
\textcolor{col_shar}{SHARK}     &   $1.3\times 10^{-7}$      &   $8.0\times 10^{-3}$      \\ \hline
\end{tabular}
\end{table}

In \noindent Figure \ref{hxrlfs} we exemplify the overall capability of these models to match the local and intermediate $z$ HXLF \citep[$2-10$\,keV, compared with observations from][]{Aird2015,Buchner2015,Miyaji2015} and RLF \citep[1.4\,GHz, compared with observations from][]{rigby11,smolcic17}. A more extensive comparison of the HXLF between predictions and observations (for some of the models presented here), covering higher redshifts (where the differences between models and observations tend to increase), can be found in \citet{Sijacki2014}, \citet{guevara2016}, \citet{volonteri16} and \citet{griffin18}.

The agreement for the local Universe is not surprising since these models have been developed to reproduce low-redshift
observations (in some cases the properties of black holes studied here). Nevertheless, it is useful to illustrate how close these models can be to the actual observations.

\subsection{Radiative efficiency ($\epsilon$)}\label{radeff}
Prior of analysing the high-redshift predictions, at X-rays and radio frequencies, from the considered models, we should address the handling of the radiative efficiency. This parameter, which denotes the efficiency of the conversion from infalling matter to energy as radiation, is commonly assumed to be constant \citep[e.g.][]{Khandai2014, Sijacki2014, guevara2016,volonteri16}. However, recent studies \citep[e.g.][]{martinez09,rong2012} suggest a possible variation of $\epsilon$ with redshift and SMBH mass. To gauge the effect of such evolution, and since variations in the values of $\epsilon$ could affect the emission from the highest redshift SMBHs, we explore now the effect of a changing radiative efficiency throughout the history of the Universe for these models.
For this study we use the GALFORM model, where the spin parameter is explicitly provided. Although, the Horizon-AGN model provides the spin parameter as well, a value of $\epsilon = 0.1$ was applied in order to avoid inconsistencies, since this value was used to run the model. We adopt the equations described in \citet{Bardeen1972} and \citet{griffin18}. The radiative efficiency is given as:
\begin{equation}
\epsilon = 1 - \sqrt{1 - \frac{2}{3} \frac{1}{\hat{r}_{\rm lso}}},
\label{first}
\end{equation}

\noindent where $\hat{r}_{\rm lso}$ is the last stable orbit of the accretion disc around the SMBH, in units of gravitational radius $R_{\rm G} = G M_{\bullet}/c^2$ and is given by:
\begin{equation}
\hat{r}_{\rm lso} = r_{\rm lso}/R_G = 3 + Z_2 \pm \left[ (3 - Z_1) (3 + Z_1 + 2Z_2)\right]^{1/2}
\label{rlso}
\end{equation}

\noindent where $Z_1$ and $Z_2$ are functions of the spin parameter $\alpha$:
\begin{gather}
Z_1 = 1 + (1 - a^2)^{1/3} [(1 + a)^{1/3} + (1 - a)^{1/3}].\\
Z_2 = (a^2 + Z_1^2)^{1/2}.
\label{last}
\end{gather}

\noindent In equation \ref{rlso} the minus sign corresponds to an orbit that has the same direction with the spin/angular momentum of the SMBH ($\alpha>0$), whereas the positive sign corresponds to a retrograde orbit ($\alpha<0$).

\begin{figure}
\centering
\includegraphics[width=\columnwidth]{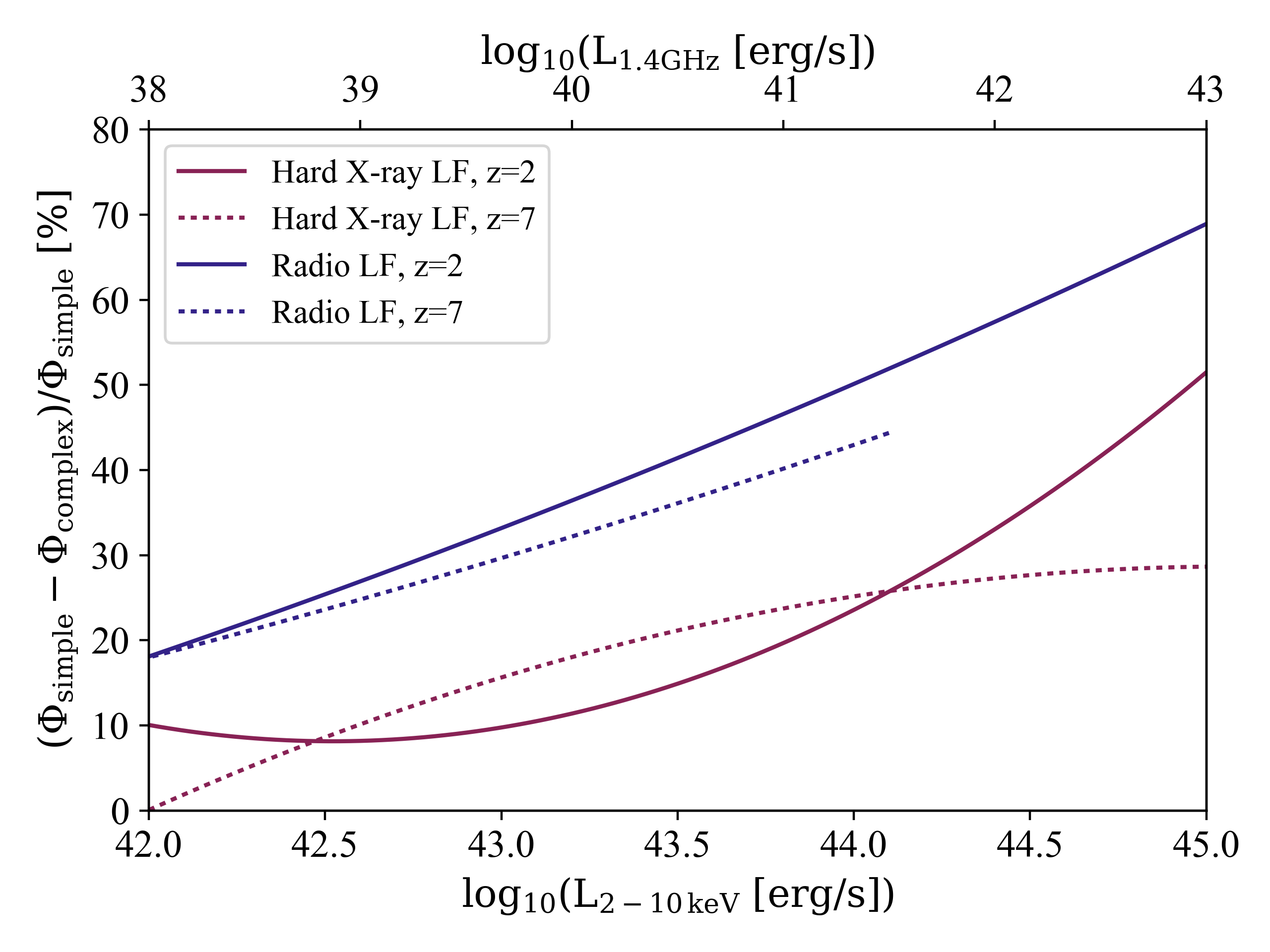}
\caption{The spline curves for the difference (in $\%$) between the simple ($\epsilon=0.1$) and a more complex calculation (equations \ref{first}-\ref{last}) of the radiative efficiency in terms of the hard X-ray (red lines) and radio (blue lines) luminosity functions for SMBHs in the GALFORM model for redshift $z=2$ (continuous lines) and $z=7$ (dashed lines). The number densities for $z=2$ and $z=7$ are $n=0.068\,\rm  Mpc^{-3}$ and $n=0.045\, \rm Mpc^{-3}$ respectively.}
\label{durham_compar}
\end{figure}

Figure \ref{durham_compar} presents the comparison between estimating the XRLF and RLF with $\epsilon=0.1$ (referred here to as simple method) and allowing it to change (referred here to as complex method).
As we see both methods produce similar results at low luminosities (less than 20(30)$\%$ difference for $L_{\rm 2-10 keV} (L_{\rm 1.4GHz}) <10^{43} (10^{40})$ erg/s). However, for the most luminous SMBHs ($L_{\rm 2-10 keV} (L_{\rm 1.4GHz}) >10^{44}(10^{41})$ erg/s) the difference in the X-ray and radio LFs can be higher than 40$\%$. This difference affects the luminosity estimates, in particular at the highest redshifts, since we expect to be able to observe the most luminous and massive SMBHs. Therefore, assuming a constant value for the efficiency might not be appropriate for the earliest epochs.

In this work we explicitly calculate the radiative efficiency and/or spin whenever this is possible (GALFORM), adopting constant values in all other situations ($\epsilon = 0.1$ for Horizon-AGN, EAGLE, MassiveBlackII, L-Galaxies and SHARK; $\epsilon = 0.05$ for Illustris; $\epsilon = 0.06$ for MERAXES), where SMBH spin and/or radiative efficiency is not explicitly handled.

\subsection{AGN in the Epoch of Re-ionization}
In order to explore the predictions of the models considered here for the highest redshifts, we focus on the redshift range $6\leq z \leq 10$, in the EoR \citep[e.g.][]{Zaroubi2012}, and generate the high-z HXLF and RLF for each model considered. Although obscuration in the hard X-rays regime is considered negligible, we apply a correction for possible obscuration effects to the X-ray emission as given in \citet{Aird2015}, considering a range of minimum and maximum obscuration values of $N_H<10^{21} \, \rm cm^{-2}$ and $N_H>10^{24} \, \rm cm^{-2}$ respectively.

\begin{gather}
\frac{d\Phi_{\rm abs}}{dlogL_{\rm x}} =
\begin{cases}
  (1-f_{\rm 21-22}) \frac{d\Phi}{dlogL_{\rm x}}, \, \, \, \, [10^{20}<N_{\rm H} \rm [cm^{-2}]<10^{21}]    \\
  \frac{\beta_{\rm cthick}}{2} \frac{d\Phi}{dlogL_{\rm x}}, \hspace{0.65cm} [10^{24}<N_{\rm H}\rm [cm^{-2}]<10^{26}]
\end{cases}
\label{obsc}
\end{gather}
where $f_{\rm 21-22}=0.43$ is the fraction of unabsorbed AGN and $\beta_{\rm cthick}=0.34$ is a normalization factor for the Compton-thick AGN. This obscuration limits will be the lower and upper errorbars in the predicted LFs.

\begin{figure*}
\centering
\includegraphics[width=\columnwidth]{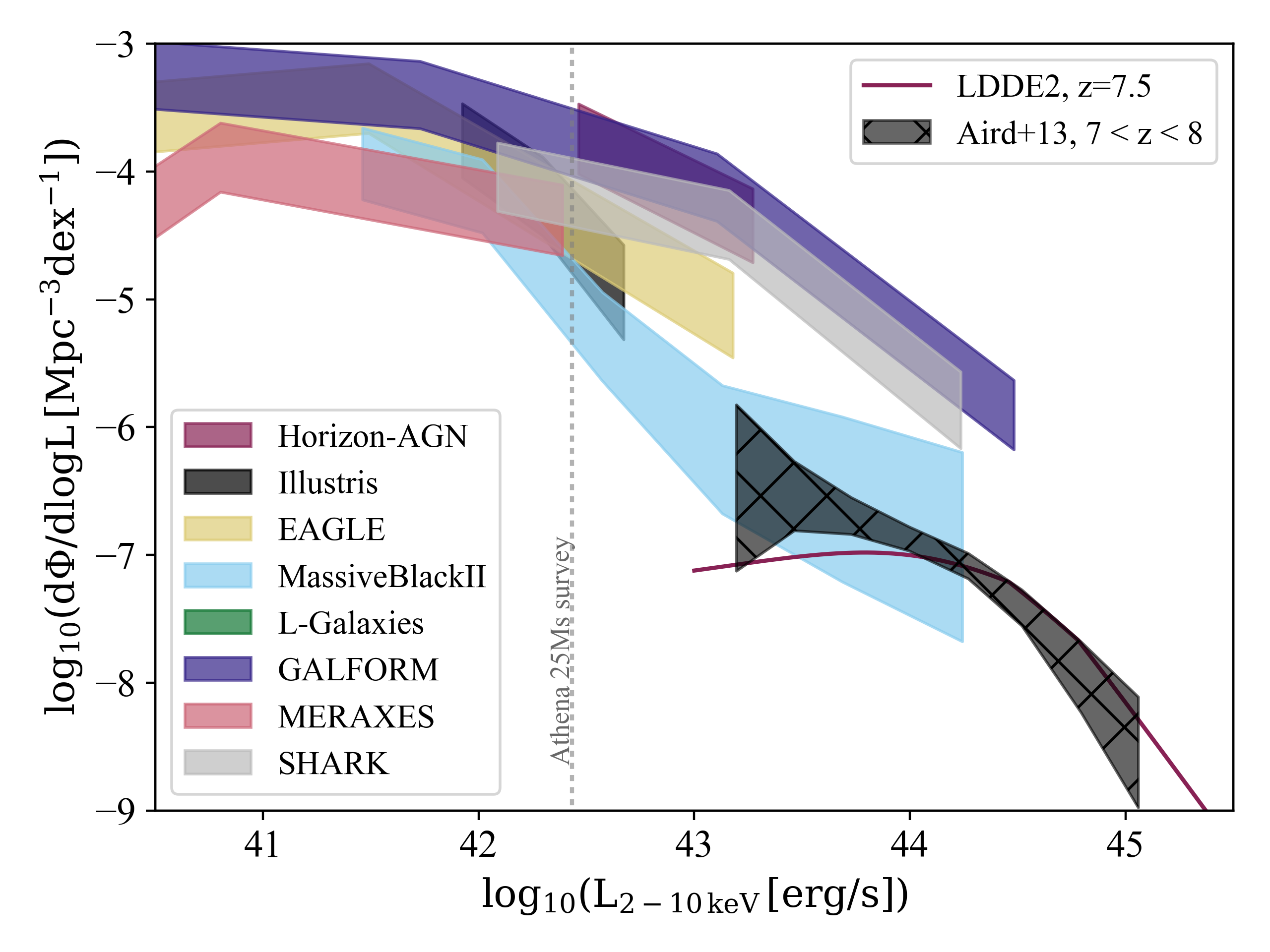}
\includegraphics[width=\columnwidth]{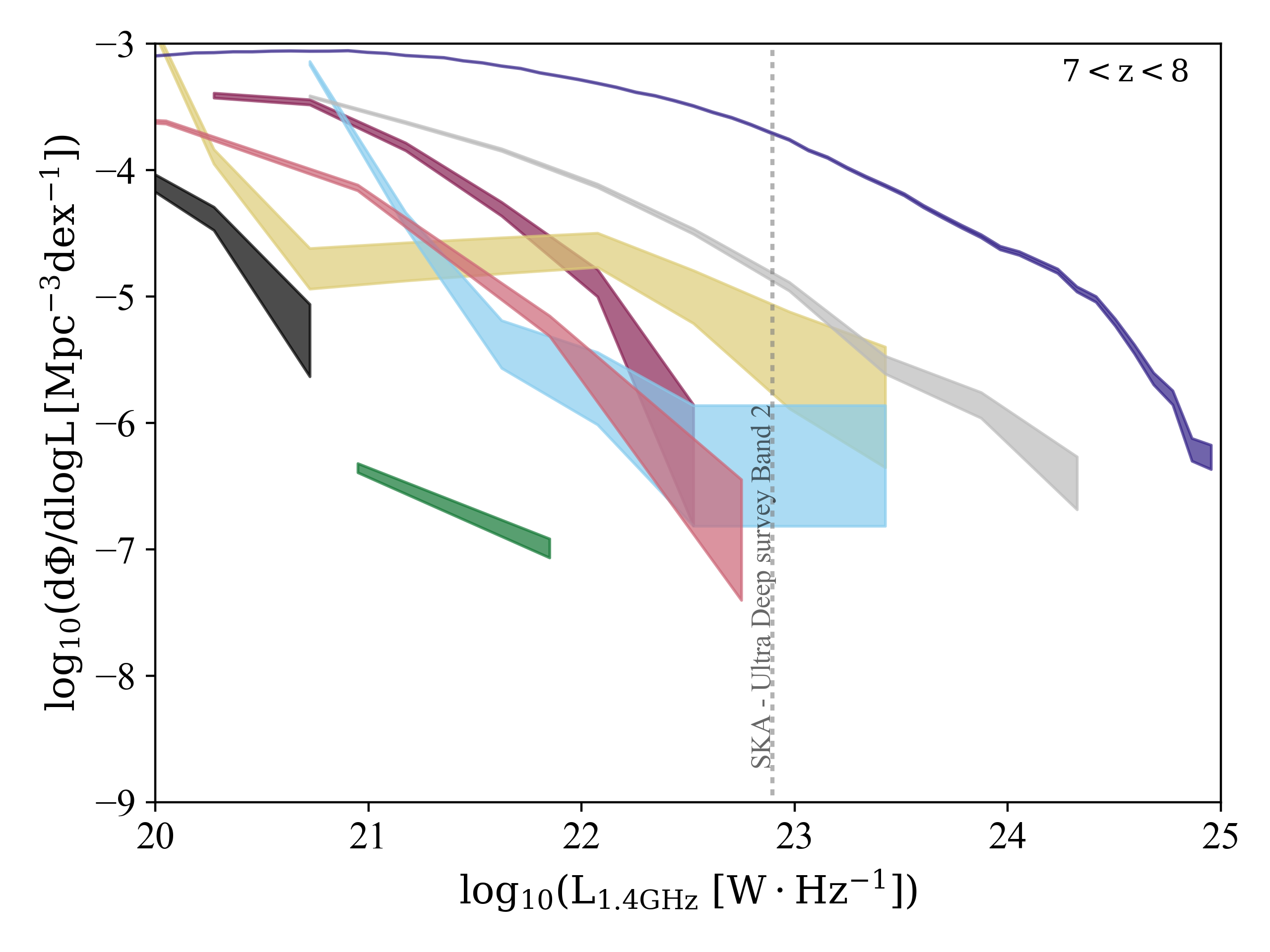}
\caption{The hard X-ray (left panel) and radio (right panel) luminosity functions for the redshift range $7<z<8$. In the left panel we also display the LDDE2 model (solid line) and the model presented in \citet{aird2010} (black hatched region). The dashed lines depict the luminosity sensitivity limits for the predicted Athena 25 Ms and SKA Ultra Deep future surveys at band 2 ($L_{\rm 2-10keV,lim}=2.7 \times 10^{42}\, \rm erg/s$ and $L_{1.4\rm GHz,lim}=7.9 \times 10^{22}\, \rm W/Hz$ respectively). The shaded area for each model in the left plot is derived by using equation \ref{obsc} estimating the minimum and maximum obscuration effect, while for the radio part it represents the Poisson error. Mass limits (similar to Figure \ref{hxrlfs}) were not applied in this calculations.}
\label{810}
\end{figure*}

The resulting estimate for the LFs after these obscuration corrections for the redshift range $7<z<8$ is presented in Figure \ref{810}. Additionally, for the HXLFs two theoretical models are also presented in the figure for comparison. The first one (solid red line) is a Luminosity-Dependent Density Evolution model \citep[LDDE2 as described in][]{Aird2015} given by:
\begin{equation}
d\Phi / dlogL_{\rm x} = K\, \bigg[\bigg(\frac{L_{\rm x}}{L_{\ast}}\bigg)^{\gamma_1}+\bigg(\frac{L_{\rm x}}{L_{\ast}}\bigg)^{\gamma_2}\bigg]^{-1} \cdot e(z, L_{\rm x}),
\end{equation}
where $L_{\ast}$ is the characteristic break luminosity, K is a normalization constant, $\gamma_{1,2}$ are the slopes of this broken power-law and $e(z, L_{\rm x})$ is the z and luminosity evolution factor.
The second model also shown is an extrapolation of the local XLF to higher redshifts \citep[][]{aird2010}.
A final correction step is applied to take into account the redshift of the emitted photons, considering an X-ray photon index $\Gamma$ of 1.4 in:
\begin{equation}
L_{\rm X} = 4\pi d_{\rm L}^2 f_{\rm X} (1+z)^{\Gamma - 2},
\label{conversion}
\end{equation}
where $L_{\rm X}$ is the X-ray luminosity, $d_{\rm L}$ is the luminosity distance of the source and $f_{\rm x}$ is the observed hard X-ray flux. In this sense, a luminosity limit of $L_{\rm X}=10^{43} \, \rm erg/s$ translates into $f_x \approx 10^{-17} \, \rm erg/s/cm^2$ and $f_x \approx 10^{-17.5} \, \rm erg/s/cm^2$ in the redshift ranges $z=7-8$ and $z=8-10$ respectively.

As far as the RLFs are concerned no obscuration was applied, however considering the redshifted photons emitted the following equation was applied \citep[][]{afonso06}:
\begin{equation}
L_{\rm 1.4 GHz} = 4\pi d_{\rm L}^2 S_{\rm 1.4 GHz} 10^{-33} (1+z)^{\Gamma - 1},
\label{conversion_radio}
\end{equation}
where $L_{\rm 1.4 GHz}$ is the RLF, $d_{\rm L}$ is the luminosity distance of the source and $S_{\rm 1.4 GHz}$ is the 1.4 GHz flux density in units of $\rm mJy$. The spectral index $\Gamma$ was set to 0.8 a typical value for synchrotron radiation.

\section{Results} \label{results}
Figure \ref{hxrlfs} shows that the model-predicted HXLFs and RLFs at low and intermediate redshifts are in reasonable agreement with observations, a result which is not surprising as these are often part of the constraints imposed during the development of the models. While this does not mean high-z's will be equally successful, it at least suggests a reasonable degree of accuracy in the physical processes considered in the models. It should thus be possible to trace the evolution of the AGN population and generate robust predictions for the observation of the highest redshift ranges, both in X-rays and radio frequencies. The use of different models can provide a measure of the uncertainties affecting the highest redshifts, as well as potentially highlighting the need for specific improvements in the models. This is complementary to many works being developed today \citep[e.g.][]{scubed, aird2010}, that explore the highest redshifts via semi-empirical simulations or some more or less complex (but always uncertain) extrapolation from intermediate redshift observations.

\subsection{Detection of AGN at the Epoch of Re-ionization}\label{results_eor}

In this work we are particularly interested in the model predictions for redshift ranges corresponding to the Epoch of Re-ionization (EoR) and the detectability of the early stages of galaxy formation through their early AGN activity. At such high redshifts ($z\sim 6-10$) we will be exploring what missions like Athena, in the X-rays, or SKA and SKA-precursors, in the radio, will potentially be able to reveal. One should realise that knowledge of the physical conditions in the early Universe can be rather incomplete \citep[consider, for example, the hard-to-quantify CMB-muting effect -- energy losses of emitting electrons by Inverse Compton to the hot CMB at very high redshifts -- that will predominantly affect extended radio emission,][]{ghisellini2014,afonso15}, but the predictions will help in guiding future radio and X-ray surveys to fine-tune strategies for the detection of these sources which will ultimately lead to robust tests and consequent improvements of the models themselves.

Following the procedures detailed in Section \ref{methods}, we have estimated the LFs (and consequently the density of SMBHs, by integrating the former) that each model predicts at $6\leq z \leq 10$ (for redshift bins of $\Delta z \sim 1$). Considering their estimated luminosity and number density, at both hard X-rays ($2-10$ keV) and radio (1.4 GHz), we have explored their detectability with Athena and SKA.
In the X-rays, at the redshift range $7<z<8$ (one of the redshift bins) we have considered a sensitivity limit of $f_{\rm 2-10keV}=1.58 \times 10^{-17} \, \rm erg/s/cm^2$ \citep[][]{aird2010}, which translates to a luminosity of $L_{\rm 2-10keV}=2.7 \times 10^{42} \, \rm erg/s$ assuming a spectral index of 1.4.
At radio wavelengths, we assumed an ultra-deep reference survey for SKA \citep[][]{ska} reaching a sensitivity level of $S_{\rm 1.4GHz}= 0.2\,\rm \mu$Jy at 1.4GHz, which corresponds to a luminosity of $L_{\rm 1.4GHz}= 7.9 \times 10^{22} \, \rm W/Hz$ (for $7<z<8$) assuming a spectral index of 0.8. As a result, the number of AGN at $6\leq z \leq 10$ in the models that have luminosities above the limiting values (which vary between different $z$ bins) of each telescope is presented in Table \ref{table1}.
These results reveal strong differences between the models, with predictions varying from zero to a few thousands SMBHs detected over a square degree. Therefore, the use of only one model when it comes to predicting the SMBH population at the EoR is highly risky.
In Figure \ref{810} we present the corresponding LFs estimates for the redshift range $7<z<8$, along with the LDDE2 and Aird's 2013 model, based on extrapolation from lower redshift LFs \citep[][]{aird2010}. The grey vertical dashed lines represent Athena's and SKA's sensitivity limits. We can see that, while most models predict a significant number of detectable AGN at the highest redshifts in the X-rays (although showing a wide range in predicted numbers), the same is not seen in the radio where most models do not reveal a substantial number of AGN able to be detected by SKA (with the exception of GALFORM). As we detail below, and although this can be the result of different physics in the models (e.g. including disk instabilities) and the lack of observations that can anchor the models at intermediate redshifts, one major effect seems to be coming from the limited volumes of the simulations, making them unable to predict the highest mass SMBHs that would in general produce the highest radio luminosities (Section \ref{methods} and Appendix \ref{appB}). This aspect is further explored in the next subsection. The study of the impact of varying SMBH seed masses on the luminosity and mass functions has been conducted for the GALFORM and MERAXES models (\citealt{griffin18} and \citealt{qin17} respectively), showing
that it is only important for relatively low mass and less luminous SMBHs. Since our predictions focus on the most massive SMBHs different seed mass should not affect our results. Nevertheless we note that even before SKA, the radio detection of very high redshift AGN can still be achieved with upcoming wide area radio surveys. For example, the Evolutionary Map of the Universe \citep[EMU,][]{norris}, to be performed with The Australian Square Kilometre Array Pathfinder (ASKAP), assuming a sensitivity limit of 10 $\rm \mu Jy$, should be able to detect a few thousand very high-redshift AGN over the full 30,000 $\rm deg^2$ covered (estimates from GALFORM and SHARK only, as the remaining models reveal no detectable sources over the simulation boxes considered).

\begin{table*}
\caption{The number of SMBHs per $\rm deg^2$ at $6 \leq z \leq 10$, and their detectability by Athena or SKA for both hard X-ray and radio regimes for the  models considered.}
\label{table1}
\hspace*{-0.7cm}
\begin{tabular}{|c|c|c|c|c|c|c|c|c|c|}
\hline
Telescope & z range & \textcolor{col_hor}{Horizon-AGN} & \textcolor{black}{Illustris} & \textcolor{col_eag}{EAGLE} & \textcolor{col_mii}{MassiveBlackII} & \textcolor{col_mun}{L-Galaxies} & \textcolor{col_dur}{GALFORM} & \textcolor{col_drag}{MERAXES} & \textcolor{col_shar}{SHARK}\\ \hline
\multicolumn{ 1}{|c|}{Athena} & $6-10$ & 16577 & 3649 & 947  & 502  & 28 & 11958 & 1545 & 6274
\\ \hline
\multicolumn{ 1}{|c|}{SKA} & $6-10$ & 14 & 0 & 87  & 24  & 0 & 3434 & 17 & 292
\\ \hline
\end{tabular}
\label{table1}
\end{table*}

\subsection{Model explorations}
\begin{figure}
\centering
\includegraphics[width=\columnwidth]{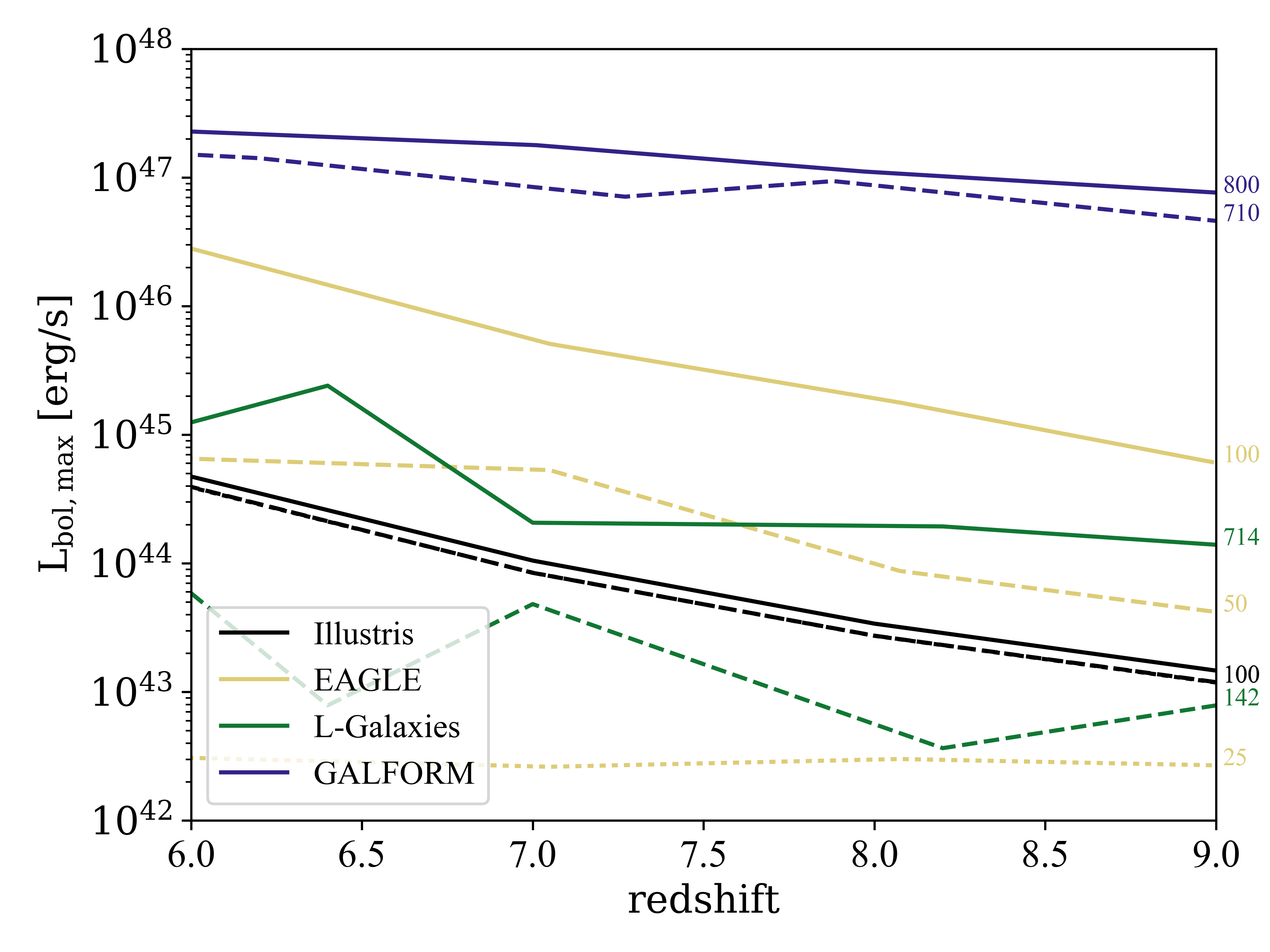}\\
\caption{The maximum bolometric luminosity for the redshift range 6-9 for the Illustris, EAGLE, L-Galaxies and GALFORM models, where more than one simulation run was used. There is a general trend for every redshift to produce more luminous AGN as the volume of the simulation increases, which can be seen for the EAGLE, L-Galaxies and GALFORM models. On the other hand with the Illustris simulation it is shown that same volume but different physical conditions produce similar results. Dotted lines present the smallest volumes for the model they follow, dashed intermediate volumes and the straight lines the largest volumes. The box size of each model can be seen in the right edge of the plot with the representative colour.}
\label{diff_simulations}
\end{figure}

\subsubsection{Volume effect}
Smaller simulation boxes appear unable to predict significant numbers of massive, and luminous AGN at high redshifts. In order to explore the effect of volume in the final high-redshift predictions, we perform a simple comparison between the different simulation runs of the Illustris, EAGLE, L-Galaxies and GALFORM models. For the EAGLE project we use 3 simulations which share the same physics and numerical techniques with box sizes of 25, 50 and 100 Mpc (with resolutions that also vary accordingly) and are denoted as Ref-25, Ref-50 and Ref-100. As far as the Illustris project is concerned, 3 different simulation runs are used, with the same box size of 106.5 Mpc, same cosmological parameters and initial conditions but with different mass and spatial resolutions. For L-Galaxies we use the two available DM simulations, Millennium 1 and 2, which differ in their volumes (714 Mpc and 142 Mpc, respectively). Finally, for the GALFORM model we have 2 runs of volume size 710 Mpc and 800 Mpc corresponding to the Millennium-I and Millennium-P DM simulations (which use different cosmological parameters and resolution). Figure \ref{diff_simulations} displays a comparison between the maximum bolometric luminosity observed in each simulation at the redshift range $z=6-9$. We can see that runs of the same models (same physics) with different volumes produce higher bolometric luminosities for increasing box volume. This is very striking when comparing the 3 EAGLE runs, for example, for which increasing the volume by a factor of 8 corresponds to a high redshift increase in the maximum bolometric luminosity that can be significantly larger than a factor of 10. The different resolutions that usually accompany the changes in volume do not appear to justify this increase, as indicated by the behaviour of the Illustris simulations — for which fixing the volume and changing resolutions alone does not seem to significantly impact the maximum bolometric luminosity (difference of less than 20$\%$).

\begin{figure*}
\centering
\includegraphics[width=\columnwidth+\columnwidth]{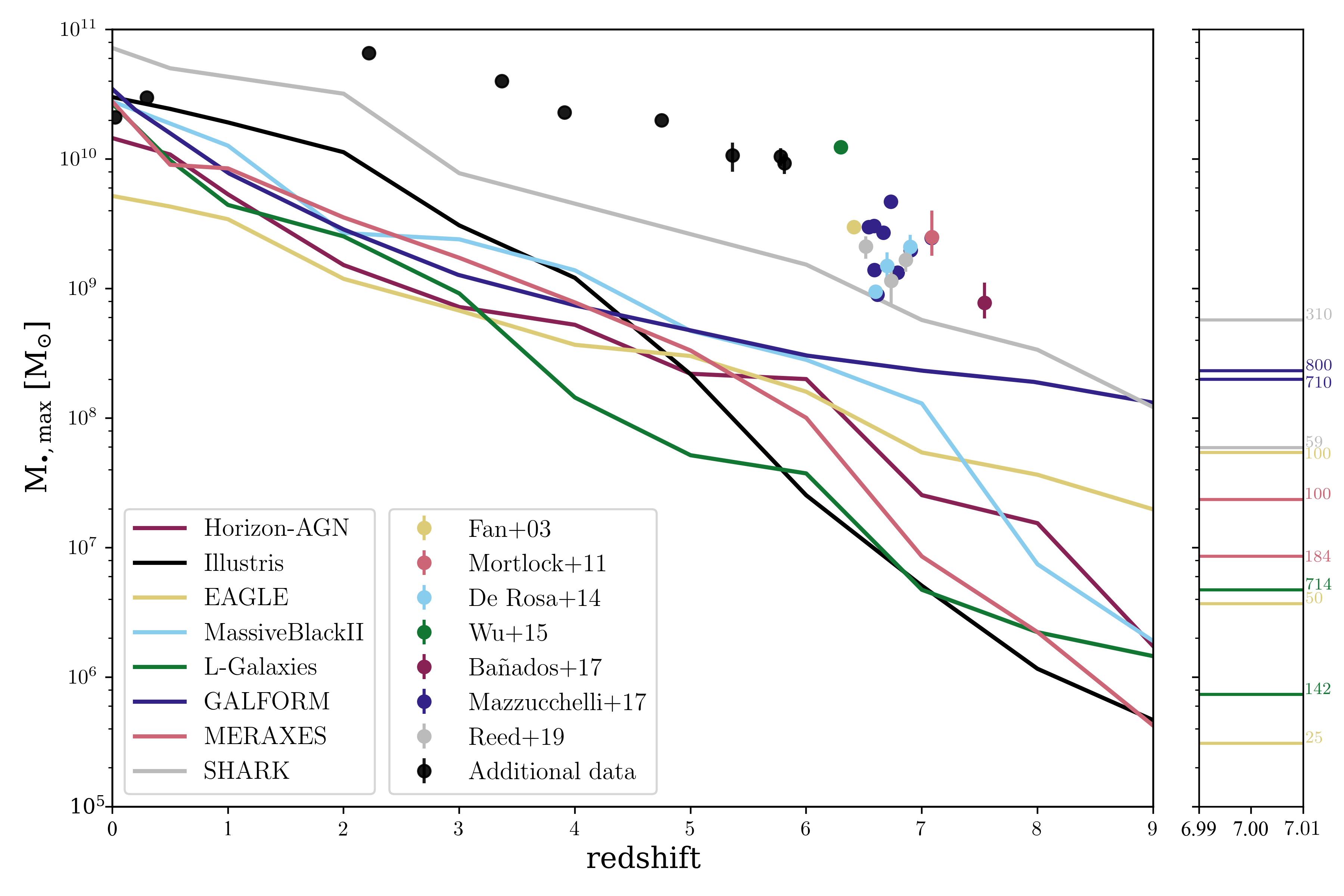}
\caption{The most massive SMBHs produced by each model for redshifts $z<9$ (solid lines). For comparison, observational data are presented (coloured data points with errorbars) for $z>6$ \citep[][]{fan,mortlock,derosa,wu,mazzu,banados,reed} and for lower redshifts (black points with errorbars) \citep[][]{shemmer04, 2009ApJ...690..463R, 2011Natur.480..215M, 2015ApJ...799..189Z, 2015PASP..127...67B, wolf18}. On the right subplot, the horizontal coloured lines represent the same model (depending on the colour) but with different volume for redshift $z=7$. The volume of each of these additional models is given at the right end of the subplot with the corresponding colour of the model. In this sense for the EAGLE model (with yellow colour) we have three data points for $z=7$ for box sizes of 25, 50 and 100 Mpc.}
\label{maxmass2}
\end{figure*}

A similar exercise can help to better understand the limitations of state-of-the-art galaxy simulations when used to predict the highest redshift Universe. In Figure \ref{maxmass2} we directly compare the maximum SMBH mass at each redshift as predicted by different models, with recent observational data of powerful quasars. The panel on the right shows the box length for each simulation (at $z=7$) in units of Mpc. Models that provide additional simulation runs with different volume, are depicted with the same colour (e.g. the SHARK model has two grey lines corresponding to the simulations of box size 59 and 310 Mpc). In general, simulations substantially under-predict the maximum SMBH masses for all but the lowest redshifts. This difference exists already at the highest observed redshifts ($z\sim 7$), revealing limitations in the early rate of SMBH growth or constraints in the production of the most extreme objects. However, it has to be noted that these observations are detections of high $z$ quasars, found in large sky surveys covering a much larger volume (at the EoR) than the models used in this work. For example, the $z=7.54$ quasar of SMBH mass of $M_{\bullet}=8\times 10^{8} \, \rm M_{\odot}$ \citep[][]{banados}, was detected in the UKIRT Infrared Deep Sky Survey \citep[UKIDSS,][]{UKIDSS} covering a sky area of 4028 $\rm deg^2$, which at the detected redshift corresponds to a box with length of $\sim 74 \, \rm Gpc$ (using a scale of 5.125 $\rm kpc/''$, $H_0=67.7 \, \rm kms^{-1}Mpc^{-1}$, $\Omega_{\rm M} =0.307$ and $\Omega_{\rm \Lambda}=0.693$). None of the models presented here can reach such a large simulation box, which might be necessary in order to create the most luminous and massive SMBHs, since there is a general trend to predict more massive SMBHs when increasing the simulation box volume (MERAXES is the only model contradicting this result for the high redshift regime, however according to \citealt{qin17} this can be seen as an effect of the low resolution/merging rate of the highest volume simulation). From the same figure we can see that smaller size models (e.g. SHARK) can create more massive SMBHs than larger volume models (e.g. L- Galaxies) which shows that beyond the volume effect the specific choices on how accretion is modelled impact the results.

In any case, Figure \ref{maxmass2} indicates that predictions from current state-of-the-art galaxy models should be taken as lower limits to the actual number of rare, high luminosity AGN at very high redshifts, and consequently the corresponding LFs. Quite importantly, these results show that simply to increase the volume of the models, which will certainly happen over
the coming years, will lead to a significant improvement when it comes to match model predictions and observations at high redshifts. A final depiction of the volume effect is presented in Appendix \ref{appB}.

\subsubsection{Accretion modes at high redshift}
\begin{figure*}
\centering
\includegraphics[width=\columnwidth]{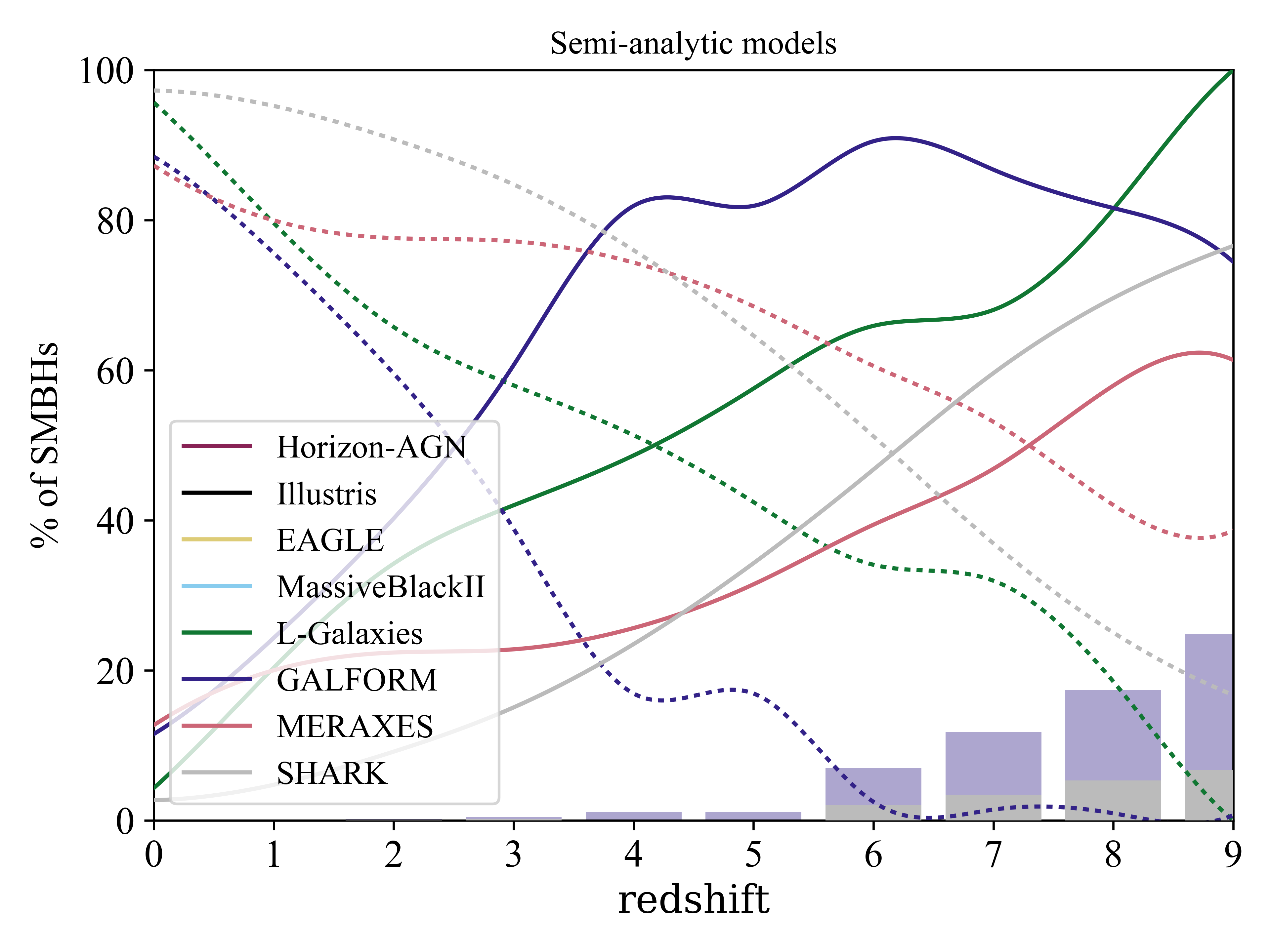}
\includegraphics[width=\columnwidth]{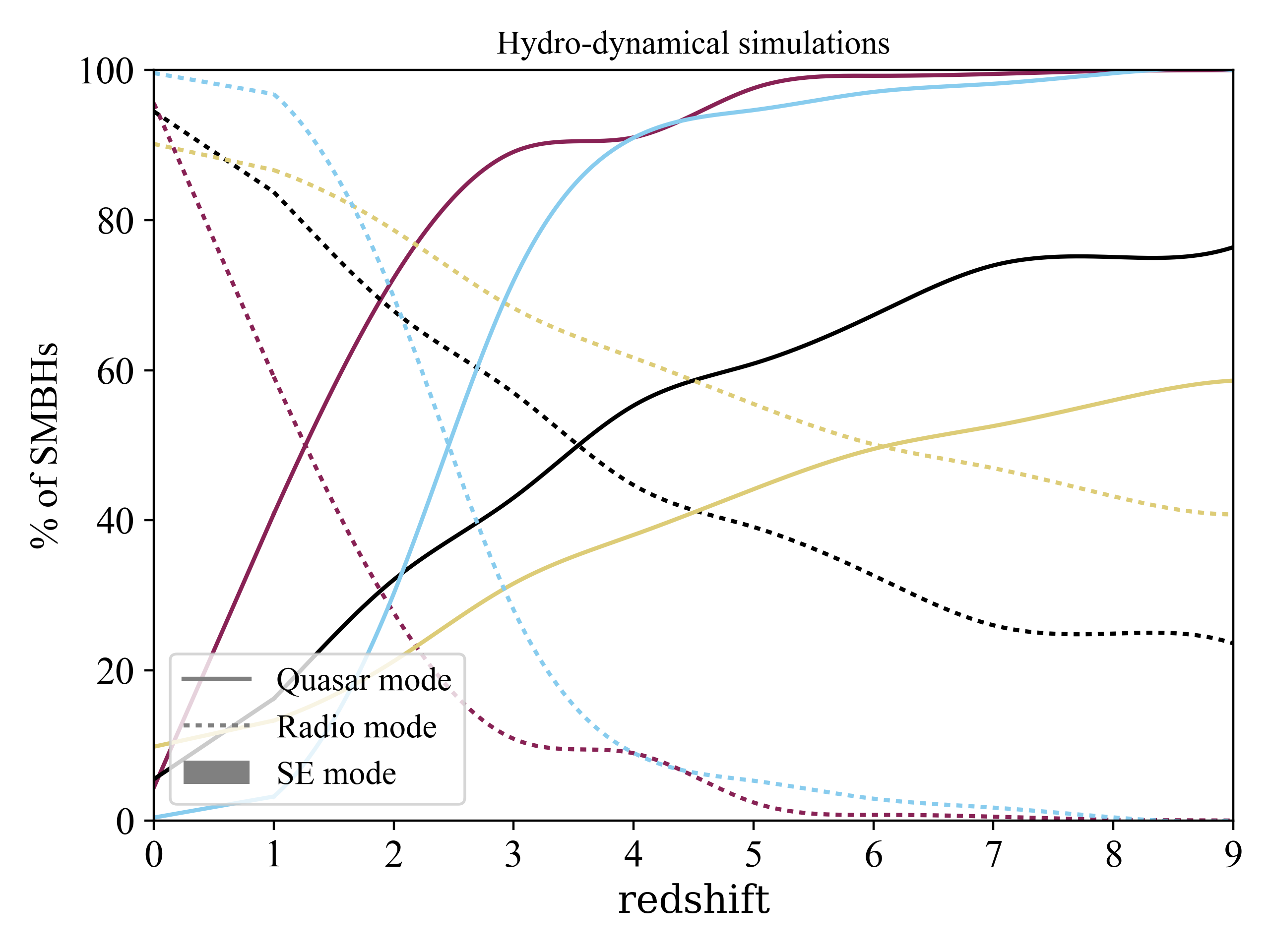}
\caption{The percentage of SMBHs growing via quasar (solid lines) and radio (dashed lines) accretion modes for each model (left plot for the semi-analytic models and right plot for the hydro-dynamical simulations) for redshifts between 0 and 9. The bars depict the percentage of SMBHs accreting at Super-Eddington rate with only GALFROM and SHARK showing values higher than 5$\%$. Only SMBHs of mass higher than $10^5 \,\rm  M_{\odot}$ and Eddington ratio of $\lambda=L_{\rm bol}/L_{\rm Edd}>10^{-6}$ were selected.}
\label{per_modes_1}
\end{figure*}

Current radio surveys fail to identify high redshift ($z > 6$) radio galaxies, in spite of the sensitivity limit being presumably deep enough to observe them. An interesting question that arises is the contribution of the two accretion modes to these high redshift radio galaxy populations. The models used in this work should be able to offer valuable insights about what we currently expect to observe at the highest redshifts. In Figure \ref{per_modes_1} we present the percentage of SMBHs that accrete through radio (dashed lines) and quasar (solid lines) modes. The percentage of Super-Eddington (SE) accreting sources is also presented (bars). All models are in agreement to a SMBH growth that mostly takes place by radio mode at low redshifts, whilst for the high redshift Universe the quasar mode dominates. Although this is a reflection of the models trying to produce the most massive galaxies very quickly at the beginning of the Universe, it is an indication that at high redshifts the radio emission from accretion to a SMBH may not be as abundant, or as easily produced, as X-rays which will be abundant from quasar mode accretion. In any case it is worth noting that radio emission is far from inexistent even for high accretion rates, as can be seen from Figure \ref{810}. Finally, the SE contribution becomes important only for $z > 6$ and only for the GALFORM and SHARK models which might suggest that in addition to the large simulation volume, this accretion mode is necessary at high z in order to predict the SMBH population we observe since the GALFORM and SHARK models are the ones that approach closest to the observations that we have at $z > 6$ (see Figure \ref{maxmass2}).

\section{Summary and Discussion}\label{discussion}
With the next generation of telescopes currently developing survey strategies with a strong emphasis in exploring the early Universe (e.g. Athena in the X-ray, SKA in the radio regime), it is of the utmost importance to explore the predictions from state-of-the-art galaxy formation and evolution models, in particular at the Epoch of Re-ionization (EoR).

With this goal in mind, we have followed the early evolution of the AGN/SMBH population in eight up-to-date and observationally tested hydro-dynamical simulations (HDSs) and semi-analytic galaxy formation and evolution models (SAMs), providing predictions to the number of AGN that Athena and SKA may reveal at $6 \leq z \leq 10$ and, as a result, to the AGN X-ray and radio luminosity functions. For the conversion from AGN source densities to luminosities one needs to assume an efficiency of the conversion of infall matter (to the SMBH) to energy -- radiative efficiency ($\epsilon$). It is common to assume a constant value of $\epsilon=0.1$ in these calculations, although there are indications in the literature that this value may vary throughout the Universe history and with SMBH mass \citep[e.g.][]{martinez09,rong2012}. Employing the output of the models, we show that the assumption of a constant efficiency will lead to significant differences in the estimate of the AGN luminosities -- in particular at the highest SMBH masses (and rarest, but also potentially more luminous). Explicitly including the spin of the SMBH in the models can help to better handle the efficiency in the luminosity estimates, something that will be necessary as more precise explorations of the EoR are made.

The model-derived local X-ray and radio AGN/SMBH luminosity functions are found to compare well to recent observational data. Considering this as a proof that the physics of galaxy formation and evolution are at least relatively well handled by the simulations, we explore the model predictions at very high redshifts ($z>6$) at X-ray and radio wavelengths. As far as the X-ray regime is concerned, although the various models differ in the prediction of the number of SMBHs that Athena will be able to detect in the future, the typical values of a few $\times 10^{3}$\,SMBHs/deg$^2$ are one order of magnitude higher than prior predictions from the Athena team \citep[][]{aird2010, athena_white}, which are based on extrapolations of the observed X-ray LFs to $z\sim 3-4$, assuming an evolution to higher redshifts. At radio wavelengths, the models suggest the detection of a lower number of AGN  (few $\times 10^{2}$\,SMBHs/deg$^2$) for SKA deep surveys, a result of the lower predominance of the more radio luminous accretion mode at the highest redshifts. However, our estimates are heavily dependent on two normalization parameters necessary to calculate the radio luminosity, A$_{\rm ADAF}$ and A$_{\rm TD}$, which present a high degree of degeneracy and cannot be estimated from first principles. In order to break this degeneracy, we find the values of these parameters that fit best the local Universe and higher redshifts. Nevertheless, it is noteworthy that, even in the situation that quasar mode accretion dominates, radio emission can still be quite substantial, leading to a potentially large number of radio detections at $z>6$. This lends support to the effort of finding radio AGN at the EoR, as a way to not only understand the earliest examples of AGN activity, but also to directly study the HI 21cm forest against a bright radio AGN placed in the EoR \citep[e.g.][]{carilli04} -- a point of extreme interest to the radio selection of very high-z AGN.

Finally, we show that both X-ray and radio LF estimates should be considered only as lower limits, as all models are still unable to reproduce the most extreme SMBH masses already known to exist at very high redshifts, an effect which seems to arise from the limited volume of the simulations. This leads to a likely significant underestimation of the number of rare, high luminosity AGN at very high redshifts and, consequently, to the derived LFs. While this effect is difficult to quantify, future increases in the simulation volumes and resolution (even above the current maximum $\sim 1$\,Gpc linear dimensions used) are still needed to approach the most extreme AGN examples already observed in the Universe. Such improvements are currently implemented in such models as for example in the recently published IllustrisTNG simulation \citet[][]{illustristng} or the T-RECS model - \citet[][]{trecs}.

It is also clear that further improvements on the physical prescriptions in the models are still needed and will have a significant impact on the final results. This is illustrated by the comparison with galaxy formation and evolution models that have been developed only for very high redshifts. Although outside of the scope of this work, restricted to models that have been tested and shown to reproduce several (low-redshift) observables, such high-redshift-only models can reveal higher maximum SMBH masses at $z>6$ than the majority of the models explored here. It is worthy of note that different approaches to the early stages of galaxy formation phenomena exist and can lead to even better models in the future, upon being tested against observations.

\section*{Acknowledgements} \label{acknowledgements}
SA, JA, HM, IM and CP gratefully acknowledge support from the Science and Technology Foundation (FCT, Portugal) through the research grants PTDC/FIS-AST/2194/2012, PTDC/FIS-AST/29245/2017, UID/FIS/04434/2013 and UID/FIS/04434/2019. CDPL is funded by an Australian Research Council Discovery Early Career Researcher Award (DE150100618) and by the Australian Research Council Centre of Excellence for All Sky Astrophysics in 3 Dimensions (ASTRO 3D), through project number CE170100013. The Galform runs used the DiRAC Data Centric system at Durham
University, operated by the Institute for Computational Cosmology on
behalf of the STFC DiRAC HPC Facility (www.dirac.ac.uk). This
equipment was funded by BIS National E-infrastructure capital grant
ST/K00042X/1, STFC capital grants ST/H008519/1 and ST/K00087X/1, STFC
DiRAC Operations grant ST/K003267/1 and Durham University. The work done using GALFORM was supported by the Science and Technology facilities Council ST/L00075X/1. AJG acknowledges an STFC studentship funded by STFC grant ST/N50404X/1. We acknowledge the support from the teams developing the models presented in this paper, for making their data available.

%%%%%%%%%%%%%%%%%%%%%%%%%%%%%%%%%%%%%%%%%%%%%%%%%%


%\bibliographystyle{mnras}
\bibliographystyle{mnras}
\bibliography{mybib}{}

%%%%%%%%%%%%%%%%%%%%%%%%%%%%%%%%%%%%%%%%%%%%%%%%%%

%%%%%%%%%%%%%%%%% APPENDICES %%%%%%%%%%%%%%%%%%%%%
\appendix
\section{Normalization parameters}\label{appA}
\begin{figure}
\centering
\includegraphics[width=\columnwidth]{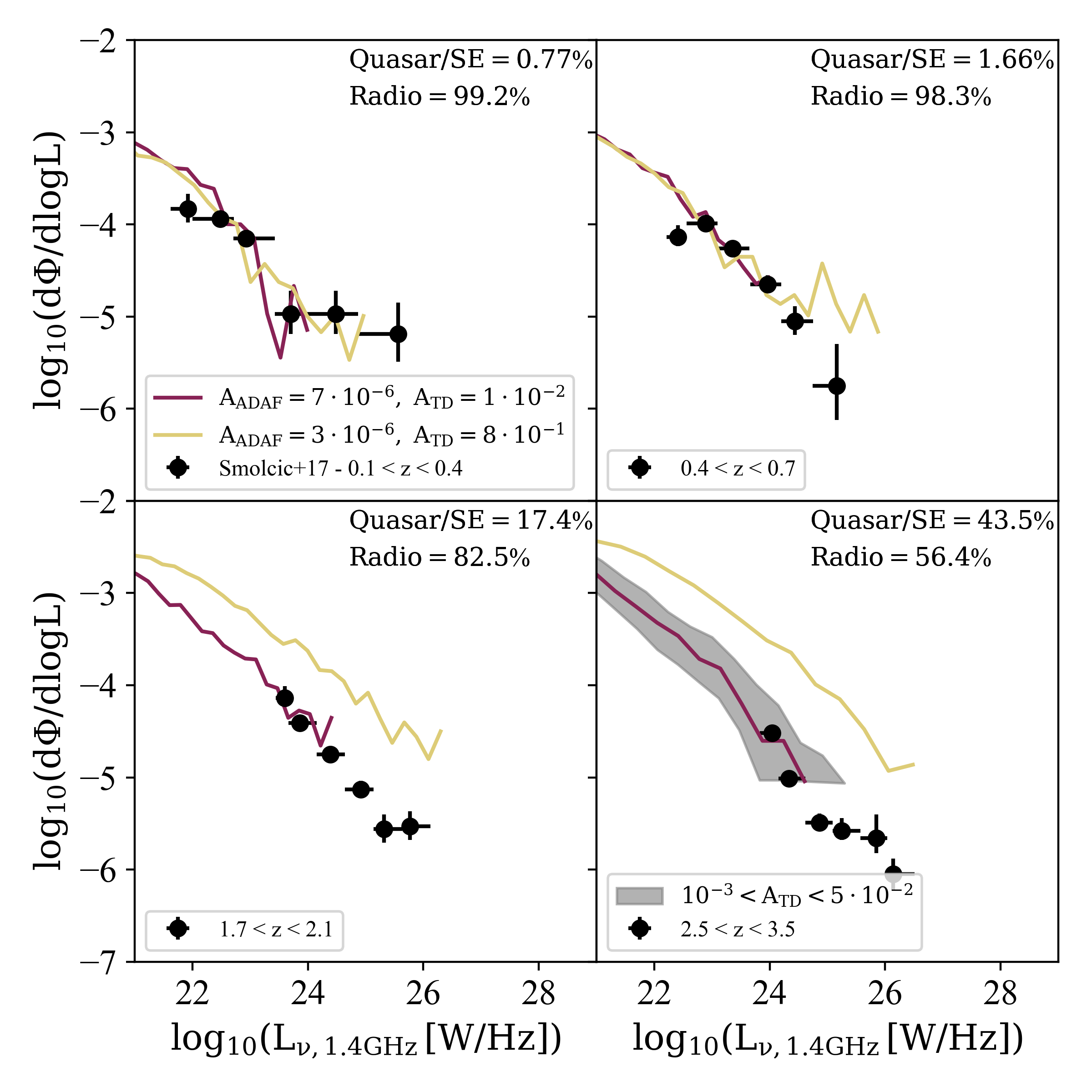}
\caption{The radio LFs (continuous lines) for 4 different redshifts using 2 different set of normalization parameters for the Illustris model. In the calculations we include all SMBHs independent of mass or rate. In order to achieve the best value for the $\chi^2$ minimisation we use data from \citet{smolcic17}. The percentage of radio and quasar (including Super Eddington accretion) mode are noted at the top of each subplot. For the last redshift range (bottom right plot) the grey shaded region shows the range of acceptable fitting by using the parameter values range which can provide a rough estimate for the error in the predictions for the number of SMBHs at the EoR detectable by SKA.}
\label{figurea1}
\label{degeneracy}
\end{figure}

As it was briefly mentioned in subsection \ref{radiolfs}, there is a high degree of degeneracy between the two normalization parameters of the total radio luminosity. As we can see in Figure \ref{degeneracy}, where the RLFs for various redshifts for the Illustris simulation are presented, this degeneracy is obvious since two different sets of parameters (red and yellow lines) provide a good fitting to the observational data for the local Universe. Since this low redshift is being used most commonly to tune the free parameters of a model one can be mislead by the aforementioned degeneracy. One of the main reasons for this issue is the low number of SMBHs accreting at quasar mode at the local universe. Consequently, the $A_{\rm TD}$ is not contributing significantly to the shape of the radio luminosity function, allowing one to reproduce similar functions with very different values of $A_{\rm TD}$. The solution we are using in this work to break this degeneracy is to provide the best fitting for various redshift ranges  using the same set of normalization parameters. In this way even though two sets of parameters can provide good fitting at low redshift one of those options (yellow line in Figure \ref{figurea1}) may not be consistent with the observational data at higher redshift. Figure \ref{degeneracy} shows two sets of those normalization parameters as well the $A_{\rm TD}$ value range at $2.5<z<3.5$ which can be perceived as an error in our predictions for the SKA future surveys. Since the quasar mode dominates the EoR, omitting a similar value range for the $A_{\rm ADAF}$ is acceptable.

\section{Volume effect}\label{appB}
\begin{figure}
\centering
\includegraphics[scale=0.5]{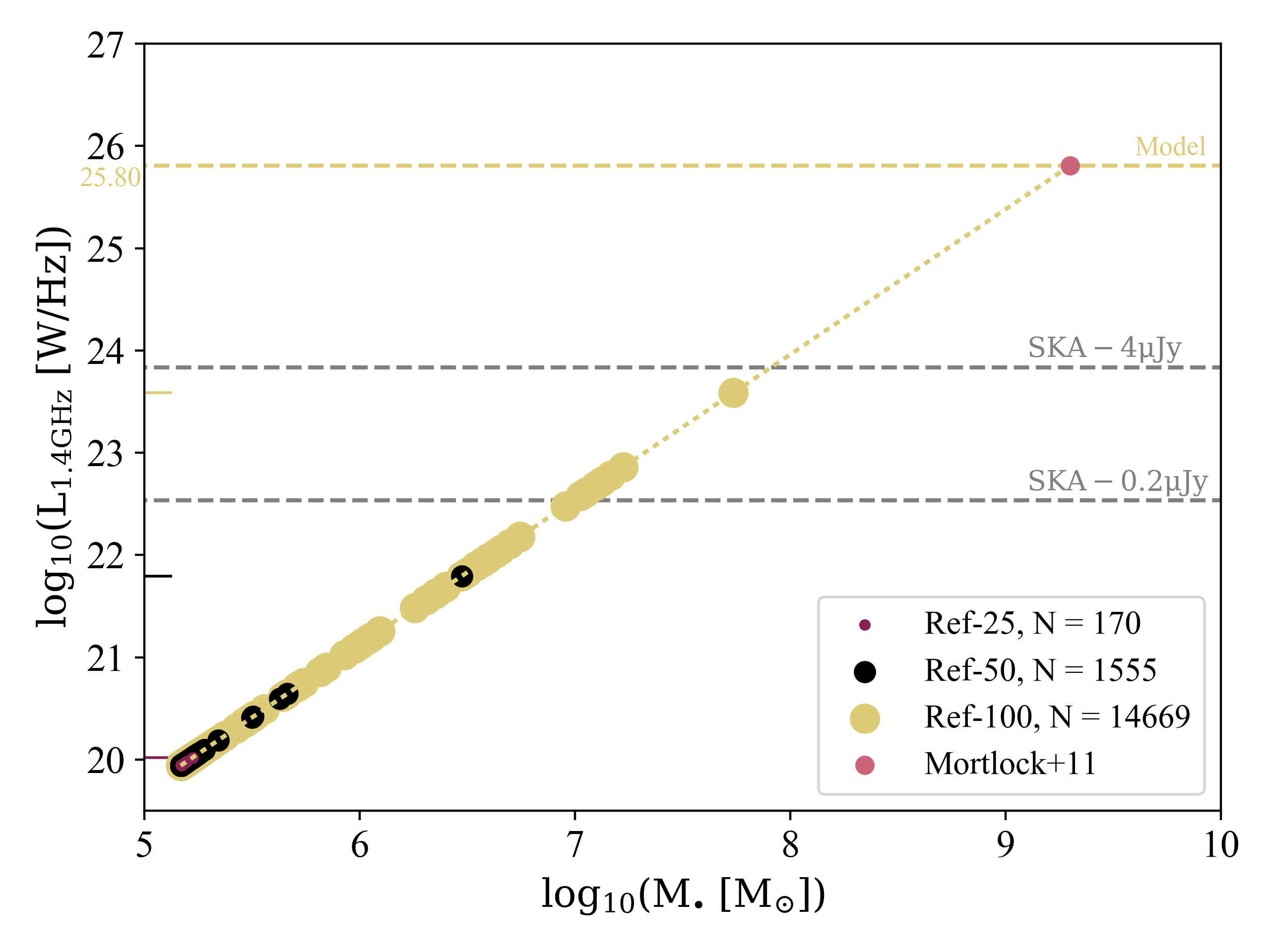}
\caption{The radio 1.4 GHz luminosity versus the SMBH mass (both in log scale) from the three available versions of the EAGLE model for $z=7$ and only for SMBHs accreting at the quasar mode. Along with the models the \citep[][]{mortlock} observation is presented as well two sensitivity limits from the band 2 SKA future survey with 4 and 2 $\rm \mu Jy$ flux limits (grey dashed lines). In the legend of the plot the number of sources for each version of the model is also presented. The coloured ticks on the luminosity axis denote the maximum luminosity that each version of the simulation can reach.}
\label{appb_fig}
\end{figure}

Another way to issue the effect of the volume of the models in the predictions for the most massive and energetic SMBHs is by looking only at the quasar mode scenario since it dominates the accretion for z>6 in all models. The motivation arises from equations 4-7 where the radio luminosity $\nu L^{\rm TD}_{\rm \nu}$ depends only on the SMBH mass:
\begin{equation}
L^{\rm TD}_{\rm \nu} \propto M_{\bullet}^{0.32} \dot{m}^{-1.2} L^{\rm TD}_{\rm jet} \propto M_{\bullet}^{0.32} \dot{m}^{-1.2} M_{\bullet}^{1.1} \dot{m}^{1.2}\propto M_{\bullet}^{1.42}
\label{eq_appb}
\end{equation}
This linear relation shows that the higher the SMBH mass accreting at quasar mode the higher the radio luminosity emitted. This view has great impact on our predictions for the SKA surveys as the limitation in maximum mass of the small volume models have been already shown. In other words if a model of small volume cannot exceed the sensitivity limit of SKA, a larger simulation box of the same model may surpass this limit, if we accept that larger volumes provide the most extreme SMBH masses. This effect can be seen in Figure \ref{appb_fig} where the EAGLE model of 3 simulation volumes is presented (for $z=7$). If we follow the yellow line (EAGLE model of 100 Mpc box size) for the SKA Band-2 $=4 \, \rm \mu Jy$ survey we get 0 SMBHs as prediction, since the maximum luminosity produced is below the SKA limit.
However, the maximum SMBH mass from this model at $z=7$ is below $10^8 \, \rm M_{\odot}$ even though observationally we know there is at least one SMBH with mass above $10^{9} \, \rm M_{\odot}$ \citep[][]{mortlock}. Since the relation between the radio emission and SMBH mass is a power-law (equation \ref{eq_appb}) an EAGLE version with a larger volume might be able to produce a SMBH with mass $\sim 10^{9} \, \rm M_{\odot}$ and whose radio emission exceeds the $4\, \rm \mu Jy$ limit.

\end{document}